\documentclass[a4paper,twocolumn,10pt,accepted=2025-07-23,longbibliography]{quantumarticle}
\pdfoutput=1
\usepackage[svgnames]{xcolor}

\usepackage[numbers,sort&compress]{natbib}
\usepackage[utf8]{inputenc}
\usepackage[T1]{fontenc}
\usepackage{amsmath,amssymb}
\usepackage{graphicx}
\usepackage{dcolumn}
\usepackage{enumitem}
\usepackage[normalem]{ulem}
\usepackage{float}
\usepackage{bm,dsfont}
\usepackage{multirow}
\usepackage[export]{adjustbox}
\usepackage{simpler-wick}
\usepackage{cuted}
\usepackage{color}
\usepackage{lipsum}
\usepackage{hyperref}
\hypersetup{
	colorlinks = true,
	linkcolor = RoyalBlue,
	citecolor = ForestGreen,
	urlcolor = RoyalBlue}

\newcommand{\cov}{\mathrm{cov}} 
\newcommand{\bra}[1]{{\langle #1|}}
\newcommand{\ket}[1]{{|#1 \rangle}}
\newcommand{\braket}[2]{{\langle #1|#2\rangle}}

\newcommand{\Ket}[1]{{|#1 \rangle\!\rangle}}

\newcommand{\dd}{\mathrm{d}}
\newcommand{\ii}{\mathrm{i}}
\newcommand{\id}{\mathds{1}}
\newcommand{\e}{\mathrm{e}}

\newcommand{\Cl}{{\mathcal{C}\ell}}

\newcommand{\scE}{\mathcal{E}}

\newcommand{\scM}{\mathcal{M}}

\newcommand{\Tr}{\mathop{\operatorname{Tr}}}

\newcommand{\eq}[1]{\begin{equation}#1\end{equation}}
\newcommand{\eqs}[1]{\begin{equation}\begin{split}#1\end{split}\end{equation}}
\newcommand{\eqnref}[1]{Eq.\,\eqref{#1}}
\newcommand{\figref}[1]{Fig.\,\ref{#1}}
\newcommand{\tabref}[1]{Tab.\,\ref{#1}}
\newcommand{\secref}[1]{Sec.\,\ref{#1}}
\newcommand{\refcite}[1]{Ref.\,\cite{#1}}

\usepackage[mathlines]{lineno}

\usepackage[capitalise,nameinlink]{cleveref}

\begin{document}

\title{Dual-unitary shadow tomography}

\author{Ahmed A. Akhtar}\thanks{a1akhtar@ucsd.edu}
\affiliation{Department of Physics, University of California San Diego, La Jolla, California 92093, USA}
\affiliation{Quantum Artificial Intelligence Laboratory (QuAIL), NASA Ames Research Center, Moffett Field, California 94035, USA}
\affiliation{KBR, Inc., 601 Jefferson St., Houston, Texas 77002, USA}

\author{Namit Anand}\thanks{namit.anand@nasa.gov}
\affiliation{Quantum Artificial Intelligence Laboratory (QuAIL), NASA Ames Research Center, Moffett Field, California 94035, USA}
\affiliation{KBR, Inc., 601 Jefferson St., Houston, Texas 77002, USA}

\author{Jeffrey Marshall}\thanks{dr.jeffrey.s.marshall@gmail.com}
\affiliation{Quantum Artificial Intelligence Laboratory (QuAIL), NASA Ames Research Center, Moffett Field, California 94035, USA}
\affiliation{USRA Research Institute for Advanced Computer Science, Mountain View, California 94043, USA}

\author{Yi-Zhuang You}\thanks{yzyou@physics.ucsd.edu}
\affiliation{Department of Physics, University of California San Diego, La Jolla, California 92093, USA}


\begin{abstract}
We introduce ``dual-unitary shadow tomography'' (DUST), a classical shadow tomography protocol based on dual-unitary brick-wall circuits. To quantify the performance of DUST, we study operator spreading and Pauli weight dynamics in one-dimensional qubit systems, evolved by random two-local dual-unitary gates arranged in a brick-wall structure, ending with a measurement layer. We do this by deriving general constraints on the Pauli weight transfer matrix and specializing to the case of dual-unitarity. Remarkably, we find that operator spreading in these circuits have a rich structure resembling that of relativistic quantum field theories, with massless chiral excitations that can decay or fuse into each other, which we call left- or right-movers. We develop a mean-field description of the Pauli weight in terms of $\rho(x,t)$, which represents the probability of having nontrivial support at site $x$ and depth $t$ starting from a fixed weight distribution. We develop an equation of state for $\rho(x,t)$ and simulate it numerically using Monte Carlo simulations. For the task of predicting operators with (nearly) full support, we show that DUST outperforms brick-wall Clifford shadows of equal depth. This advantage is further pronounced for small system sizes and our results are generally robust to finite-size effects.
\end{abstract}

\maketitle

\tableofcontents

\section{Introduction}
\label{sec:intro}

The dynamics of quantum information in closed, thermalizing systems is a core to our understanding of quantum many-body physics \cite{Deutsch1991,Srednicki1994,Rigol2008}. In thermalizing quantum systems, local quantum information is scrambled between constituent components and the entanglement exhibits volume-law scaling. Initially localized operators spread ballistically and their evolution across space and time can be used to define various measures of information propagation, such as the out-of-time-order (OTOC) correlator \cite{Larkin1969QuasiclassicalMI,Lieb:1972wy,D_ra_2017,Nahum_2017,Nahum_2018,von_Keyserlingk_2018,brown2013scrambling,styliaris_2021_scrambling,zanardi_2021_openOTOC}. The physics of these systems is often contrasted with the phenomena of many-body localization, where a closed, interacting many-body quantum system can evade thermalization, and local quantum information can escape global scrambling \cite{Nandkishore_2015}. Rather than studying localization, however, our focus in this paper is restricted to thermalizing systems generated by dual-unitary circuits. Dual-unitary circuits are fast-scrambling circuits that produce chaotic dynamics and furthermore have a symmetry that enables efficient calculation of observables \cite{Borsi_2022}. 

Our primary object of study is the \textit{Pauli weight} of an operator. The Pauli weight of a fixed operator $O$ characterizes how much of it is distributed over different Pauli strings: $w_O(P)=\Tr(OP)^2$, where $P=\bigotimes_i P_i$ is a Pauli string i.e. a product of local Pauli operators $P_i\in \{ \id,X,Y,Z \}.$ By studying the evolution of the Pauli weight over time and over different Pauli strings, we can analyze how quantum information spreads over space and time, similar to previous works on the operator spreading in random unitary circuits \cite{Ho2017E1508.03784,Bohrdt2017S1612.02434,Nahum2017Q1608.06950,Kukuljan2017W1701.09147,Nahum2018O1705.08975,von-Keyserlingk2018O1705.08910,Khemani2018O1710.09835,Rakovszky2018D1710.09827,Chan2018S1712.06836,Zhou2019E1804.09737,Zhou2019O1805.09307,Qi2019M1906.00524,Xu2019L1805.05376,Chen2019Q1808.09812,Parker2019A1812.08657,Kuo2020M1910.11351,Akhtar2020M2006.08797,von-Keyserlingk2022O2111.09904,Schuster2022O2208.12272}. The Pauli weight is also interesting from the perspective of classical shadow tomography \cite{Huang2020P2002.08953, Ohliger2013E1204.5735, Guta2018F1809.11162,Huang2021E2103.07510, Hadfield2020M2006.15788, Elben2020M2007.06305, Enshan-Koh2022C2011.11580, Hu2022H2102.10132, Hu2023C2107.04817, Levy2021C2110.02965, Bu2022C2202.03272, Hu2022L2203.07263, Seif2022S2203.07309, Hao-Low2022C2208.08964, Akhtar2023S2209.02093, Bertoni2022S2209.12924, Arienzo2022C2211.09835, Ippoliti2023O2212.11963, Hu2024D2402.17911}, where it may be used to characterize the sample complexity of predicting the expectation value of a Pauli string $P$ \cite{Hu2023C2107.04817,Akhtar2023S2209.02093,akhtar2023measurementinduced}. In this work, we will use our observations about the Pauli weight in dual-unitary circuits to analyze their usefulness for predicting extensive properties of quantum states. 

The primary scientific question we aim to address is how locally-scrambled random dual-unitary circuits differ from conventional random Haar brick-wall circuits in their Pauli weight dynamics, and what this implies about the advantages or disadvantages of using dual-unitaries for prediction. By thinking of the Pauli weight as a state vector in a many-body quantum system, we are able to interpret the dynamics in the familiar language of condensed matter physics. We learn that dual-unitary Pauli weight dynamics are unique in that they allow for the existence of stable, chiral excitations which we call left and right-movers, in analogy with the components of the Dirac field. These chiral excitations are freely propagating and are responsible for the ballistic front in the average weight. The chiral movers form the left and right branches of a future light cone when studying the Pauli weight of contiguous operators evolved by random dual-unitary circuits. These chiral movers, and therefore the front of the average Pauli weight, propagate at the ``speed of light''  ($v=1$) and contribute to the fast-scrambling of quantum information in the system (see for e.g., Refs. \cite{Bertini2020,holdendye2023fundamental} which also discuss chiral excitations in dual-unitary circuits). The rapid rate of thermalization in random dual-unitary circuits makes them superior for predicting extensive operators compared to brick-wall arrangements of Clifford gates.

This is established by evaluating the scaling base of the shadow norm, $\beta$, for extensive operators in dual-unitary brick-wall circuits compared to Clifford brick-wall circuits, where $\beta=(\|P\|^2)^{1/n}$, and $P$ is a Pauli string operator with support on all $n$ sites. We verify that this $\beta$ can be lower in the dual-unitary case for sufficiently large scrambling parameter $\alpha$ using both numerics and analytical arguments based off of a mean-field approximation. The argument starts by defining a simple order parameter in the statistical model of chiral movers described above: \textit{the average occupation} $\rho(x,t)$ at each point in space-time $(x,t)$. The average occupation characterizes the probability of finding a particle at a given space-time point. We analyze the dynamics of this order parameter via a mean-field approximation. With our simplified model of the dynamics, it's clear that the shadow norm of contiguous operators has contributions from two competing effects in the average occupation: saturation, which drives $\rho(x,t)$ from 1 to its equilibrium value of 3/4 and thereby reduces $\beta$; and dispersion, which expands the spatial region where $\rho(x,t)$ is non-zero and thereby increases the shadow norm. Dual-unitaries are able to saturate quickly, but will generally have higher shadow norm for local operators because of their fast dispersion. We are able to eliminate the effect of dispersion by considering extensive operators, which have full or nearly full support, and achieve a lower total shadow norm compared to brick-wall Clifford circuits.

\begin{figure}[!t]
    \centering
    \includegraphics[width=0.49\columnwidth]{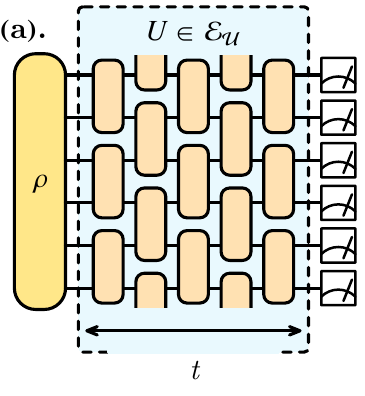} 
    \includegraphics[width=0.49\columnwidth]{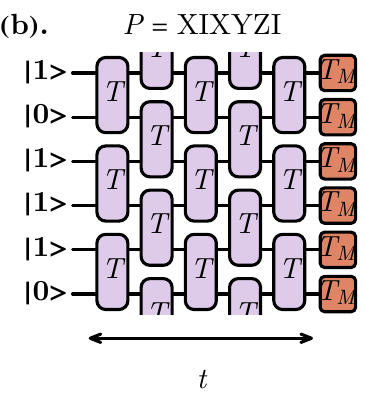} 
    \caption{(a). The traditional setup for classical shadow tomography. To perform prediction on the quantum state $\rho$, we evolve by random unitary $U$ chosen from the ensemble $\mathcal{E}_U$. At the end, we perform a final layer of measurements and store the outcomes. In post-processing, we use the classical shadows $\hat\rho$ as least-squares estimators of the state $\rho$ evaluated on certain observables. (b). The circuit that evaluates the Pauli weight $w_{\mathcal{E}_\sigma}(P)$ given the support of the Pauli string $P$. The middle section of brick-wall transfer matrices $T$ (purple) matches the structure of $\mathcal{E}_U$. Each transfer matrix $T$ defines a local Markov process, so that the resulting ``state'' before the final measurement defines a probability distribution over possible supports of the Pauli string. See \eqnref{eq:gen form} for the general form of the transfer matrix. The final layer of measurement gates $T_M$ (red) projects the resulting Pauli weight distribution onto  $\ket{0}+\frac{1}{3}\ket{1}$ where-ever a measurement was performed on the quantum state. }
    \label{fig:cst}
\end{figure}

We begin by giving a brief overview of classical shadow tomography as it is relevant to this paper in \cref{subsubsec:general-intro}. Next, we introduce the concept of dual-unitarity in circuits and our notion of random, locally-scrambled dual-unitary gates in \cref{subsubsec:dual-unitary-intro}. Equipped with this knowledge, we motivate the setup and the Pauli weight transfer matrix which defines how the Pauli weight evolves as a state vector in \cref{subsec:pauli-transfer}. Using the definitions of dual-unitary and some information-theoretic bounds, we derive the general form of the transfer matrix in the dual-unitary case in \cref{subsec:dual-unitary-transfer-matrix}. Then, we study the dynamics implied by this transfer matrix as a set of Feynman rules on the left and right-movers. Essentially, the movers are allowed to fuse and by emitted by one another, but one cannot arbitrarily transform into the other. This is the key difference with transfer matrices which are derived form two-designs as discussed in \cref{subsec:pauli-weight-dynamics}. We then introduce the average occupation $\rho(x,t)$, which characterizes the average Pauli weight on a single site, and we study its structure via Monte Carlo simulations in \cref{subsec:occupation-distribution}. Then, we derive a mean-field equation for the dynamics of the average occupation, argue for why this description remains accurate, analyze the equation of state, and solve it in the homogeneous case in \cref{subsec:mean-field}. Lastly, we argue how these dual-unitary brick-wall circuits are advantageous compared to Clifford brick-wall circuits for the prediction of extensive observables using classical shadow tomography in \cref{subsec:shadow-norm}. This establishes our dual-unitary shadow tomography (DUST) protocol.

\section{Theory}
\subsection{Classical Shadow Tomography}

\subsubsection{General Formalism}
\label{subsubsec:general-intro}

Traditionally, the interest in classical shadow tomography has been mostly geared towards experimental applications: predicting linear observables of generic states on quantum devices. However, in this work, we wish to use the general setup of classical shadow tomography as a backdrop to studying operator spreading in dual-unitary circuits. There are two primary reasons we take this approach: one, this scheme is quite general and allows us to adjust the circuit depth $t$, the system size $n$, the type of gate applied, etc.; two, the growth of the Pauli weight, which characterizes the spatial distribution of an operator, is closely related to the measurement channel of the classical shadow scheme; third, operator spreading in this context characterizes the sample complexity of predicting said operator with respect to the particular circuit model via the shadow norm. Hence, in this section, we will introduce the relevant concepts of classical shadow tomography.

Classical shadow tomography was introduced as an efficient protocol for predicting quantum state observables based on few measurements \cite{Huang2020P2002.08953}. The scheme works by generating efficient classical representations (or ``snapshots'') of the original state, usually stabilizer states, by repeatedly applying a random unitary to the original state and measuring all the qubits afterward. The specific random unitary ensemble that is used (e.g. $\Cl(2)^{\otimes n}, \Cl(2^n),$ etc.) may be suited for certain types of observables over others. 

The protocol consists of two steps: randomized measurement and classical post-processing, as illustrated in \figref{fig:cst}(a). Starting with the initial state $\rho$, apply a randomly sampled unitary $U\in\mathcal{E}_U$ so that $\rho\rightarrow U\rho U^\dagger$. Then perform a projective measurement on the transformed state $U\rho U^\dagger$ in the computational basis ($Z$-basis on each qubit independently). The resulting bit-string state $\ket{b}$ (labeled by the bit-string $b$ of measurement outcomes), as well as the performed unitary transformation $U$, will be recorded. These two pieces of information define a \textit{snapshot state} $\hat{\sigma}= U^\dagger \ket{b}\bra{b} U$. The randomized measurement protocol can be formulated as a quantum channel $\mathcal{M}$, which maps the initial state $\rho$ to the average snapshot state $\sigma$ by

\begin{equation}\label{eq: def M}
	\sigma=\mathcal{M}(\rho):=2^n\mathop{\mathbb{E}}_{\hat{\sigma}\in\scE_\sigma} \hat{\sigma}\Tr(\hat{\sigma}\rho),
\end{equation}
where $\mathcal{E}_\sigma=\{U^\dagger\ket{b}\bra{b}U\;|\;b\in\{0,1\}^{\times n}, U\in\mathcal{E}_U\}$ denotes the \emph{prior} ensemble of all possible classical snapshots. In the prior snapshot ensemble, the joint probability to sample a $(b,U)$ pair is $p(b,U)=2^{-n}p(U)$, where $p(U)$ is the probability to sample $U$ and $2^{-n}$ is the probability to sample $b$ uniformly (independent of $\rho$). It's easy to verify the above map is indeed a quantum channel. 

Furthermore, if this measurement channel $\mathcal{M}$ is tomographically complete, i.e.~distinct states $\rho$ can be distinguished by different distributions of measurement outcomes $(b,U)$, then there is an inverse map $\mathcal{M}^{-1}$ called the \textit{reconstruction map} such that $\rho=\mathcal{M}^{-1}(\sigma)$. The image of a given snapshot state $\hat{\sigma}$ under the reconstruction map is a \textit{classical shadow} $\hat{\rho}=\mathcal{M}^{-1}(\hat{\sigma})$, which provides a single-shot classical representation of the initial quantum state $\rho$. The initial state $\rho$ can be restored as the ensemble expectation of classical shadows 
\eq{\label{eq: reconstruction}\rho=\mathbb{E}_{\hat{\sigma}\in\scE_{\sigma|\rho}}\mathcal{M}^{-1}(\hat{\sigma}),} 
where $\scE_{\sigma|\rho}=\{\hat{\sigma}\sim p(\hat{\sigma}|\rho)\}$ is called the \textit{posterior} ensemble of classical snapshots, in which the snapshot states $\hat{\sigma}=U^\dagger \ket{b}\bra{b}U$ are drawn from the posterior distribution $p(\hat{\sigma}|\rho)=\bra{b}U\rho U^\dagger\ket{b}p(U)$ (i.e., the probability to sample the snapshot state $\hat{\sigma}$ by the randomized measurement of the initial state $\rho$). The reconstruction map $\scM^{-1}$ in \eqnref{eq: reconstruction} enables us to predict physical observables of the initial state based on the classical snapshots collected from randomized measurements. For example, the expectation value of any observable $O$ can be reconstructed as $\langle O \rangle = \mathop{\mathbb{E}}_{\hat{\sigma}\in\scE_{\sigma|\rho}}\Tr(O\mathcal{M}^{-1}(\hat{\sigma})).$

If the unitary ensemble $\mathcal{E}_U$ is local-basis invariant (or equivalently, Pauli-invariant), then the eigen-operators of the measurement channel $\mathcal{M}$ are Pauli strings $P$, with eigenvalue $w_{\mathcal{E}_\sigma}$ denoted as the \textit{Pauli weight} \cite{Bu2022C2202.03272,Hu2023C2107.04817,Akhtar2023S2209.02093,akhtar2023measurementinduced}. Furthermore, this Pauli weight is not dependent on the specific Pauli string but on its support. It is named as such because it characterizes the ``weight'' of an operator on a particular sub-region defined by the support of the Pauli,
\begin{equation}\label{eq:def PW}
    w_{\mathcal{E}_\sigma}(P):=\mathbb{E}_{\sigma\in\mathcal{E}_\sigma} (\Tr P \sigma )^2.
\end{equation}
This is the primary object we will be studying to characterize operator spreading. Its inverse, 
\begin{equation}
    \| P \|_{\mathcal{E}_\sigma}^2=w_{\mathcal{E}_\sigma}(P)^{-1},
\end{equation}
is the \textit{locally-scrambled shadow norm} $\| P \|_{\mathcal{E}_\sigma}^2$, which characterizes the sample complexity of predicting $P$ using the tomography scheme defined by $\mathcal{E}_\sigma$ \cite{Bu2022C2202.03272,Hu2023C2107.04817, Akhtar2023S2209.02093}.

\subsubsection{Dual-Unitarity in Qubits}
\label{subsubsec:dual-unitary-intro}

\begin{figure}[H]
    \centering
     \includegraphics[width=0.9\columnwidth]{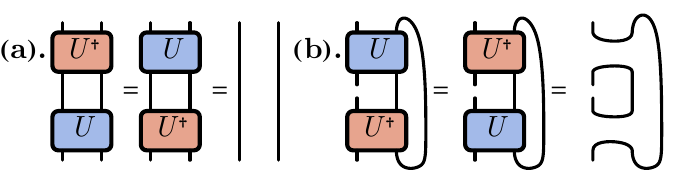}
    \caption{(a). The defining property of unitarity i.e. $U^{\dagger}U=UU^{\dagger}=\mathbb{I}.$ We denote this as \textit{temporal}-unitarity, because the indices are contracted in the time direction. (b). One of the defining properties of right spatial unitarity. There is an equivalent expression where we contract the left legs instead. Note that here we are contracting legs in a spatial direction.}
    \label{fig:dualUnitary}
\end{figure}

We are interested in simulating the evolution of an initial Pauli string $P$ under a brick-wall configuration of random, dual-unitary gates as depicted in \figref{fig:cst}. To define the ensemble average and the second moment of random, dual-unitary gates, we must first define what a dual-unitary gate is. Dual-unitary circuits have generated a lot of buzz because they're local; they're fast scramblers; they can generate chaotic dynamics; they can be used to form $k$-designs; and, they have simple, tractable expressions for two- and four-point correlation functions, including out-of-time-ordered correlators (OTOCs) \cite{Borsi_2022}.

Temporal (normal) unitary says that $U^\dagger U=\id$ so that contracting the legs in the time-directions gives back the identity. Contrarily, spatial-unitarity means that if we contract legs in the spatial direction we get back the identity (in the spatial direction). \figref{fig:dualUnitary} depicts the defining properties of dual-unitaries using tensor-network diagrams.

Consider a $n$-qubit system arranged on a lattice with Hilbert space $\mathcal{H}=\otimes_i \mathcal{H}_i$, where $\dim\mathcal{H}_i=2$. Every two-local dual-unitary qubit gate $U$ can be parameterized (up to an overall phase) by four arbitrary single-qubit unitaries $u_{\pm},v_{\pm}$, and a single coefficient $J$, which tunes the coupling in a $XXZ$-type interaction Hamiltonian \cite{PhysRevX.9.021033,PhysRevLett.123.210601,PhysRevB.101.094304,PhysRevLett.126.100603,Kos_Styliaris_2023circuitsofspacetime}. This is given below,
\begin{align}
    U(J)&=(u_{+}\otimes u_{-})V(J)(v_{-}\otimes v_{+}),\\&= \includegraphics[width=0.2\columnwidth, valign=c]{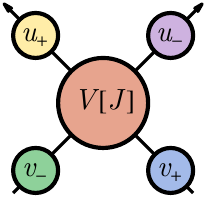},
\end{align}
where $V(J)=\exp(-i (\frac{\pi}{4}(XX+YY)+J\, ZZ))$. In matrix form, 
\begin{equation}
    V(J)= \begin{pmatrix}
        e^{-iJ} & 0 & 0 & 0 \\
        0 & 0 & -ie^{iJ} & 0 \\
        0 & -ie^{iJ} & 0 & 0 \\
        0 & 0 & 0 & e^{-iJ}
    \end{pmatrix}.
\end{equation}
At $J=\pi/4$, $V$ is proportional to a swap gate and simply interchanges sites, scrambling no information. As we turn $J$ away from $\pi/4$, we scramble more and more information until we reach $J=0$ which is akin to Hamiltonian generated scrambling with quantum $XY$-model. Additionally, the \(J=0\) point is special since it is simply an iSWAP gate, which is known to be the fastest local scrambler \cite{Bensa2021FastestScrambler}. As we will notice, this gate also generates the optimal dual-unitary shadow tomography scheme.

We are interested in the average, basis-independent behavior of dual-unitary circuits for prediction in a classical shadow setting. If we choose $u_{\pm},v_{\pm}$ according to a random unitary ensemble, then we are able to define a random, dual-unitary gate ensemble $\mathcal{E}_U$. 
\begin{equation}
    \mathcal{E}_U=\{ U(J) \sim p(U)=\dd u_{+}\dd u_{-}\dd v_{+}\dd v_{-},\}
\end{equation}
where $\dd u_{\pm}, \dd v_{\pm}$ stands for the Haar measure (i.e., uniform distribution) of the single-qubit gates $u_{\pm},v_{\pm}$.
Since this definition of the random gates is local-basis invariant, we can utilize the results of the previous section and assume that the measurement channel $\mathcal{M}$ is diagonal in the Pauli basis and the eigenvalues are given by the Pauli weight. Furthermore, we can assume that the shadow norm is a good estimate of the sample complexity. This ensemble depends crucially on $J$, and gives us a way to explore how adjusting the dual-unitary parameter affects its tomographic properties.

In practice, it suffices to sample the local scramblers $u_\pm,v_\pm$ according to the single-qubit Clifford group $\Cl(2)$. This is because $\Cl(2)$ forms a unitary 3-design over qubits, and this ensures Pauli-invariance, which in turn implies that the measurement channel is diagonal in the basis of Pauli strings \cite{Bu2022C2202.03272}.

\subsection{Storing snapshot states}
It is worth emphasizing that a key aspect of shadow tomography is the ability to store the snapshot states efficiently on a classical computer, for classical postprocessing and reconstructing the actual quantum state later. For the case of both global and shallow Clifford shadows, this follows from the Gottesman-Knill theorem, which allows us to store states of the form \(U|b\rangle\) where \(U\) is an \(n\)-qubit Clifford unitary and \(|b\rangle\) is a computational basis state (corresponding to the bit string \(b\)). However, this method does not generalize to other classes of quantum circuits. Instead, for the case of dual-unitary shadows, this follows from the fact that we only ever need to use log-depth quantum circuits for shadow tomography. Namely, for predicting an operator with full-support, i.e., \(\mathrm{supp}(O) = n\), we need a circuit depth $\sim \log(n)$. Such log-depth quantum circuits are simulable using standard matrix product state (MPS) methods \cite{Cirac2021} since we only need to simulate their evolution on computational basis states, \(U|b\rangle\). In fact, this behavior is expected more generally for any locally-scrambled random quantum circuit, including Clifford shadows \cite{Ippoliti2023O2212.11963}. We discuss the optimality of log-depth quantum circuits for shadow tomography in \cref{subsec:shadow-norm}.

\subsubsection{Set up}\label{sec:setup}
In this paper, we wish to study operator spreading in a brick-wall circuit of depth $t$ on $n$ qubits, where each gate is a locally-scrambled dual-unitary gate with parameter $J$. We will do this primarily by studying the evolution of the Pauli weight, which we will build constructively as follows. We may decompose the whole unitary $U$ applied over a single snapshot as 
\begin{equation}
    U=\prod_{i=0}^{t-1} U_{i}, \quad U_i =  \begin{cases}
        U_e & i\text{ is even} \\
        U_o & i\text{ is odd} 
    \end{cases}
\end{equation}
where $U_e$ is the even layer of the brick-wall (acting on pairs of adjacent sites starting at the first site) and $U_o$ is the odd layer (acting on pairs of adjacent sites starting at the second site), with the boundary conditions left open or closed. Using the definition of the Pauli weight \eqnref{eq:def PW}, we get that
\begin{equation}
        w_{\mathcal{E}_\sigma}(P)=2^{-n}\sum_{b\in (\mathbb{F}_2)^n}\int dU \bra{b}^{\otimes 2} U^{\otimes 2} P^{\otimes 2} (U^\dagger)^{\otimes 2} \ket{b}^{\otimes 2}.
\end{equation}

We may simplify the above expression and develop a procedure to calculate it iteratively using the following observations:

\begin{enumerate}
    \item Assume the unitary ensemble $\mathcal{E}_U$ over which we are integrating is left local-basis invariant. Then we may replace $U\rightarrow VU$ and while leaving the measure invariant, where $V=\otimes_i V_i$. If we choose $V$ such that $V\ket{b}=\ket{0}$, i.e. $V=\prod_i X^{b_i}$, we see that the Pauli weight is independent of the particular measurement outcome $b$. Hence,
    \begin{multline}
     \quad \quad \quad w_{\mathcal{E}_\sigma}(P)=  \\  \bra{0}^{\otimes 2n} \left( \int dU \, U^{\otimes 2} P^{\otimes 2} (U^\dagger)^{\otimes 2} \right) \ket{0}^{\otimes 2n}.
    \end{multline}
    
    \item Consider the action of integrating over a single gate, $U_i$, on $P$. For the Pauli weight, we are interested in evaluating the second moment, 
    \begin{equation}\int dU_i U_i^{\otimes 2} P^{\otimes 2} (U_i^\dagger)^{\otimes 2},\end{equation} 
    which we can expand in terms of Pauli operators on the doubled Hilbert space. The coefficient is
    \begin{equation}\label{eq: coeff P1P2}
        \int dU_i \,\Tr\left(  U_i^{\otimes 2} P^{\otimes 2} (U_i^\dagger)^{\otimes 2} P_1\otimes P_2 \right).
    \end{equation}
    As long as $P_1\neq P_2$, there exists a local basis rotation $V$  to flip $P_1$ as $VP_1 V^\dagger=- P_1$ (namely that \(P_1\) and \(V\) anticommute) while keeping $P_2$ invariant as $VP_2 V^\dagger=P_2$. For example, for qubit Pauli's, \(V = P_2\) fulfills this condition since any two different Pauli's anticommute. Similarly, for $k$-qubit Paulis, pick $V$ to match $P_2$ on a site where $P_1$ and $P_2$ differ, and be the identity operator on all the other sites. Under such rotation $V$, the tensor combination $P_1\otimes P_2$ in \eqnref{eq: coeff P1P2} will change sign (unless $P_1=P_2$ where $V$ cannot be found).
    Since the integration measure is invariant under any local basis rotations, the integration has to vanish for any $P_1\neq P_2$ due to its ability to change sign under local basis rotations, hence we must have $P_1=P_2$ to get a non-vanishing result for the integration. Therefore, integrating over each gate simply produces a sum over more doubled Pauli strings:
    \begin{multline}
        \quad \quad \int dU_i \,   U_i^{\otimes 2} P^{\otimes 2} (U_i^\dagger)^{\otimes 2} = \\ \sum_{P'\in \mathcal{P}_k}T^{(2)}_{P',P}( P')^{\otimes 2},\quad\quad
    \end{multline}
    where $k$ is the locality of the gate. The above matrix $T^{(2)}$ defines a Markov process over the space of Pauli strings of length $k$, which we can see by summing over the rows. The matrix elements are given by
    \begin{multline}\label{eq:def T}
        \quad\quad\quad T^{(2)}_{P',P} = 2^{-2k} \int dU_i \, (\Tr U_i P U_i^\dagger P')^2.
    \end{multline}
    Summing over $P'$ and using Fierz' identity, we get
    \begin{equation}\label{eq:norm1}
         \sum_{P\in\mathcal{P}_k}T^{(2)}_{P',P} = \sum_{P'\in\mathcal{P}_k}T^{(2)}_{P',P} = 1.
    \end{equation}
    If we continue this process layer by layer, we see that integrating over the full unitary ensemble amounts to summing over the transfer matrices $T^{(2)}_{P,P'}$ as a brick-wall tensor network. For example, if $T^{(2)}_i$ is the transfer matrix for the $i$-th layer, the Pauli weight reduces to
    \begin{multline}\label{eq:w(P) sum}
        \quad\quad\quad w_{\mathcal{E}_\sigma}(P)=\\\sum_{P_0, ... P_t} \left( \prod_{i=t}^1 (T^{(2)}_i)_{P_i,P_{i-1}}\right)\delta_{P_0=P}\delta_{P_t\in\mathcal{P}_Z}
    \end{multline}
    where the final overlap with $\ket{0}^{\otimes 2n}$ at the end constrains $P_t$ to be a $Z$-string.
    \item The above equation can be computed as a $n$-qudit tensor network over a Hilbert space with local dimension $4$. However, the Pauli weight in the locally-scrambled case only depends on the support of $P$. This can be seen from the transfer matrix elements \eqnref{eq:def T}. We can always perform a left local-basis rotation to transform $P'$ to a $Z$-string with the same support. Likewise, we can do a right local-basis rotation to transform $P$ to an equivalent $Z$-string. Since every layer is built from local-basis invariant transformations, the final result cannot depend on the specific Pauli string but only its support. This allows us to reduce the local dimension from four to two.
\end{enumerate}
Since the Pauli weight, and likewise every transfer matrix, only depends on the support of Pauli strings, we should be able to calculate $w_{\mathcal{E}_\sigma}$ as a tensor contraction on a $2^n$ dimensional Pauli weight Hilbert space $\mathcal{H}_{PW}$. We can organize the Pauli weight into a vector $\ket{w_{P,t}}$ where the weight on region $A$ is given by 
\begin{equation}\label{eq:bit-region-mapping}
    w_{P,t}(A):=\braket{b}{w_{P,t}}, \quad b_i=\begin{cases} 0 & i\notin A \\ 1 & i\in A\end{cases}.
\end{equation}
The Pauli weight vector component $w_{P,t}(A)$ is related to the Pauli weight $w_{\mathcal{E}_\sigma}(P)$ of the corresponding classical shadow protocol by 
\begin{equation}
    w_{\mathcal{E}_\sigma}(P)=\sum_{A\subset \Omega} (1/3)^{|A|}w_{P,t}(A).
\end{equation}
where $\Omega=\{1\dots n\}$. To get the Pauli weight vector for a depth $t$ brick-wall circuit, we calculate this vector's evolution through the circuit starting with an initial configuration specified by $P$. We outline the procedure below:
\begin{enumerate}
    \item Start with an initial state specifying the support of $P$: 
    \begin{equation}
        \ket{w_{P,0}} = \otimes_{i=1}^{n} \ket{\delta_{i\in \text{supp }P}}
    \end{equation}
    This is simply a computational basis state $\ket{b}$ where $b$ is a bit-string specifying the support of the Pauli string. Therefore, note that because of local-basis invariance, the initial state actually does not depend on the specific Pauli string.
    \item Define a modified transfer matrix for the Pauli weight Hilbert space $\hat T^{(2)}_{b,b'}$, where $b,b'$ are bit-strings. Let $P=\prod_i Z^{b_i},P'=\prod_i Z^{b_i'}$, then 
    \begin{equation}\label{eq:def T1}
        \hat T^{(2)}_{b,b'} = T^{(2)}_{P,P'}.
    \end{equation}
    This matrix has the same values as the transfer matrix defined in \eqnref{eq:def T}; however, it is defined on a smaller Hilbert space and so it is $4\times 4$ instead of $16 \times 16$. In order to preserve the sum in \eqnref{eq:w(P) sum}, intermediate summation variables must be weighted with the number of Paulis with non-trivial support. For an intermediate region covering $|b|$ qubits, the number of non-trivial Pauli strings with support on $b$ is $3^{|b|}$. Therefore, the full transfer matrix becomes 
    \begin{equation}\label{eq:def T2}
        \hat T_{b,b'} = 3^{|b|} T^{(2)}_{P,P'}.
    \end{equation}
    where $|b|=\sum_i b_i$ is the size of the support of $P$. 

    Given the transfer matrix $\hat{T}$, we apply it in a brick-wall fashion mirroring the geometry of the physical unitary ensemble $\mathcal{E}_U$ (see \figref{fig:cst}b). Let $\hat{T}_e$, $\hat{T}_o$ be even and odd layers of Pauli weight transfer matrices, respectively. Then, we determine the Pauli weight distribution at time $t$ as
    \begin{equation}\label{eq:wPdyn}
        \ket{w_{P,t}}=\Big(\prod_{i=0}^{t-1} \hat{T}_i\Big) \ket{w_{P,0}},
    \end{equation}
    where 
    \begin{equation}
        \hat{T}_i=\begin{cases}
        \hat{T}_e & i\text{ is even} \\
        \hat{T}_o & i\text{ is odd}. 
    \end{cases}
    \end{equation}
    \item The only Pauli strings that survive the summation in \eqnref{eq:w(P) sum} are $Z$-strings. Therefore, to evaluate the full Pauli weight, we must sum over all possible supports, but weight each non-empty region by $3^{-|b|}$, since that is the fraction of Pauli strings which are $Z$-strings. Therefore, where-ever there is a measurement gate, contract with $(1,1/3)^T$. In a standard classical shadow scheme, we end with measurements on every site, so that 
    \begin{equation}\label{eq:wP}
        w_{\mathcal{E}_\sigma}(P)=\left(\prod_{i=1}^n \left(\bra{0}+\bra{1}/3 \right) \right)\ket{w_{P,t}}.
    \end{equation}
\end{enumerate}
Now that we have a general algorithm for computing the Pauli weight vector, $\ket{w_{P,t}}$, and thereby the Pauli weight distribution and shadow norm, we proceed by deriving general conditions on the form of the Pauli weight transfer matrix and then specializing to the dual-unitary case. 

\subsection{Pauli Weight Transfer Matrix}
\label{subsec:pauli-transfer}

\begin{figure}[H]
    \centering
    \includegraphics[width=0.6\columnwidth]{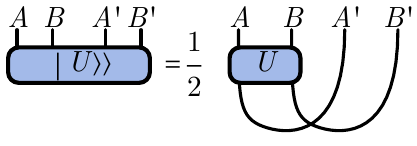}
    \caption{The Choi state representation $|U\rangle\!\rangle$ of the operator $U$. By interpreting the output legs of the unitary $U$ as input legs, we get a state on a larger Hilbert space. Then, $S^{(2)}_{|U\rangle\!\rangle}(AA')$ and $S^{(2)}_{|U\rangle\!\rangle}(AB')$ in \eqnref{eq:opE} is simply the state second Renyi entropy of $|U\rangle\!\rangle$ of the regions $AA'$ and $AB'$, respectively. }
    \label{fig:choi}
\end{figure}

Using \eqnref{eq:def T2}, we can define the Pauli weight transfer matrix in $\mathcal{H}_{PW}$ in the case of two-qubit gates as
\begin{equation}\label{eq:def T3}
    \hat{T}_{b,b'}:=\frac{3^{|b|}}{16} \int dU_i \, (\Tr U_i Z_1^{b_1} Z_2^{b_2} U_i^\dagger Z_1^{b_1'} Z_2^{b_2'})^2,
\end{equation}
where $b_i,b_j' \in \{0,1\}$ and $U$ are integrated over some yet unspecified unitary ensemble that is local-basis invariant. Therefore, the choice of $Z$ is irrelevant and any other Pauli operator would give the same result. With minimal assumptions, we can ascertain the general form of the transfer matrix $\hat{T}_{b,b'}$,
\begin{equation}\label{eq:gen form}
    \hat{T}_{b,b'}=
    \begin{pmatrix}
     1 & 0 & 0 & 0 \\
     0 & \mathcal{I}_1 &  \mathcal{I}_2 & \frac{1-\mathcal{I}_1-\mathcal{I}_2}{3} \\
     0 & \mathcal{I}_2 & \mathcal{I}_1 & \frac{1-\mathcal{I}_1-\mathcal{I}_2}{3} \\
     0 & 1-\mathcal{I}_1-\mathcal{I}_2 & 1-\mathcal{I}_1-\mathcal{I}_2 & \frac{1+2(\mathcal{I}_1+\mathcal{I}_2)}{3}
    \end{pmatrix}_{b,b'}.
\end{equation}

\begin{enumerate}
    \item \textbf{Unitarity:} From \eqnref{eq:def T3}, it's clear that $\hat{T}_{00,00}=1$. Furthermore, by unitarity and cyclicity of trace, we see that  $\hat{T}_{00,b}=\hat{T}_{b,00}=0$. Lastly, from \eqnref{eq:norm1}, the columns must sum to one. 
    \item \textbf{Time-reversal symmetry:} If the unitary ensemble is statistically time reversal invariant, such that the probability of sampling $U$ is the same as sampling $U^\dagger$, i.e. $p(U)=p(U^\dagger)$, then $\hat{T}_{01,10}=\hat{T}_{10,01}$, $\hat{T}_{11,10}=3\hat{T}_{10,11}$, $\hat{T}_{11,01}=3\hat{T}_{01,11}$. 
    \item \textbf{Spatially unbiased:} If the unitary ensemble is spatially unbiased then $\hat{T}_{10,10}=\hat{T}_{01,01}$.
    Putting all of these together gives \eqnref{eq:gen form}.
\end{enumerate}

\begin{figure}[!t]
    \centering
    \includegraphics[width=0.75\columnwidth]{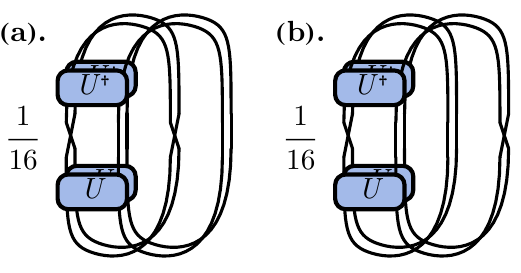}
    \caption{(a). The operator purity for the $AA'$ region i.e. $e^{-S^{(2)}_{|U\rangle\!\rangle}(AA')}$. (b). The operator purity for the $AB'$ region i.e. $e^{-S^{(2)}_{|U\rangle\!\rangle}(AB')}$.}
    \label{fig:op-purity}
\end{figure}

The remaining two independent transfer matrix elements $\mathcal{I}_1,\mathcal{I}_2$
 parameterize the operator entanglement properties of the unitary ensemble \cite{zanardi_opent_2001} via the Fierz' identity,
\begin{equation}
    2\,\mathrm{SWAP} = \id^{\otimes 2}+X^{\otimes 2}+Y^{\otimes 2}+Z^{\otimes 2}.
\end{equation}
Using the Fierz' identity, we can show that $\mathcal{I}_1$ relates to the average operator entanglement of the unitary ensemble between the $AA'$ and $BB'$ legs, whereas $\mathcal{I}_2$ is related to the entanglement between $AB'$ and $A'B$, as 
\eqs{\label{eq:opE}
    1+3\mathcal{I}_1 &= 4 \int e^{-S^{(2)}_{|U\rangle\!\rangle}(AA')}dU, \\
    1+3\mathcal{I}_2 &= 4 \int e^{-S^{(2)}_{|U\rangle\!\rangle}(AB')} dU,}
where $e^{-S^{(2)}_{|U\rangle\!\rangle}}$ is the operator purity, where \(\Ket{U}\) is the Choi state corresponding to the unitary \(U\). See \figref{fig:choi} and \figref{fig:op-purity} for a diagrammatic representation of these quantities $\mathcal{I}_{1},\mathcal{I}_2$. In terms of the Choi-state representation of $U$, 
\begin{multline}
    e^{-S^{(2)}_{|U\rangle\!\rangle}(AA')} \\=\langle\!\langle U |^{\otimes 2} \mathrm{SWAP}_A \otimes \mathbb{I}_B \otimes \mathrm{SWAP}_{A'} \otimes \mathbb{I}_{B'} | U \rangle\!\rangle^{\otimes 2},  
\end{multline}
\begin{multline}
    e^{-S^{(2)}_{|U\rangle\!\rangle}(AB')} \\=\langle\!\langle U |^{\otimes 2} \mathrm{SWAP}_A \otimes \mathbb{I}_B \otimes \mathbb{I}_{A'} \otimes \mathrm{SWAP}_{B'} | U \rangle\!\rangle^{\otimes 2}.  
\end{multline}
Since the operator entanglement can be at most one and at least 1/4, the naive bounds are $0 \leq \mathcal{I}_1,\mathcal{I}_2\leq 1$. This is already expected since the transfer matrix elements $\hat{T}_{b,b'}$ in \eqnref{eq:gen form} must all be non-negative (as conditional probabilities), which requires both $\mathcal{I}_1,\mathcal{I}_2$ to be non-negative with their total less than one. 

\begin{figure}[!t]
\begin{center}
\includegraphics[width=0.95\columnwidth]{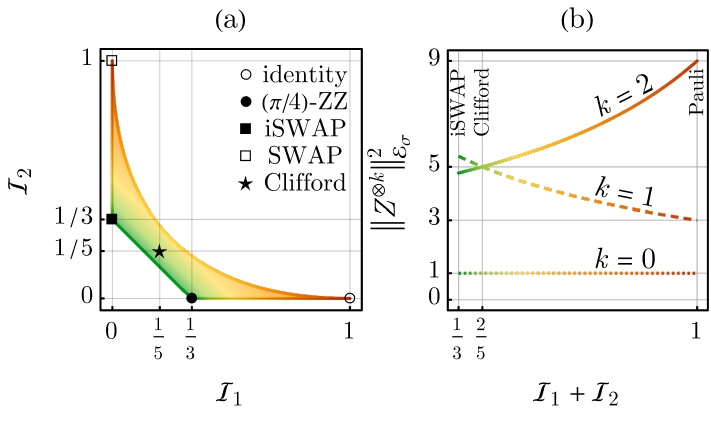}
\caption{(a) Feasible domain of $(\mathcal{I}_1,\mathcal{I}_2)$ (color by the value of $\mathcal{I}_1+\mathcal{I}_2$), with exemplary two-qubit unitary gates such as the identity gate (or any local unitary): $\id$; $(\pi/4)$-ZZ gate: $\e^{\ii\pi/4 ZZ}$; iSWAP gate: $\e^{\ii\pi/4(XX+YY)}$; SWAP gate: $\e^{\ii\pi/4(XX+YY+ZZ)}$, and random Clifford gates. (b) Shadow norms of Pauli operators of different sizes $k$ as functions of the property $\mathcal{I}_1+\mathcal{I}_2$ of the random unitary gate ensemble used in a two-qubit classical shadow tomography. The randomized Pauli measurements ($\mathcal{I}_1+\mathcal{I}_2=1$), randomized Clifford measurements ($\mathcal{I}_1+\mathcal{I}_2=2/5$), and the Pauli-twirled iSWAP measurements ($\mathcal{I}_1+\mathcal{I}_2=1/3$, a case of randomized dual-unitary measurements) are compared in the figure.}
\label{fig: gate}
\end{center}
\end{figure}

For any two-qubit unitary gate or any two-qubit random unitary ensemble, the parameters $\mathcal{I}_1,\mathcal{I}_2$ can be defined via \eqnref{eq:opE} (with or without ensemble average). Their feasible domain has been studied in \refcite{You2018E1803.10425}. We simply state the result here: the parameters must be within the box $(\mathcal{I}_1,\mathcal{I}_2)\in [0,1]^{\times2}$ and between the following bounds:
\eqs{\label{eq:I1I2bounds}
\text{lower bound: }&\mathcal{I}_1+\mathcal{I}_2\geq 1/3,\\
\text{upper bound: }&\sqrt{\mathcal{I}_1}+\sqrt{\mathcal{I}_2}\leq 1,}
and both bounds are tight (meaning that there exist two-qubit unitary gates along these boundaries). The feasible domain is shown in \figref{fig: gate}(a). Interestingly, some well-known two-qubit Clifford gates (and their Pauli-twirled versions) happen to locate at the corners of the region. The random Clifford (or Haar) ensemble corresponds to $\mathcal{I}_1=\mathcal{I}_2=1/5$, which sits inside the feasible domain. It is worth emphasizing that these bounds can also be understood by replacing the range of $\mathcal{I}_{1}$, $\mathcal{I}_{2}$ with those of the entangling power and gate typicality, respectively, as discussed in Refs. \cite{PhysRevLett.125.070501,PhysRevA.109.022610}. The general form of the transfer matrix in \eqnref{eq:gen form} was also proposed in \refcite{2024arXiv241104898K}, where the explicit dependence of $\mathcal{I}_{1}$, $\mathcal{I}_{2}$ on gate parameters is given. 

For the purpose of improving the sample efficiency of classical shadow tomography, our goal is to search for random gates that scramble as fast as possible to reduce the shadow norm. To this end, we consider a two-qubit classical shadow tomography scheme, with random unitary gate ensembles parameterized by $(\mathcal{I}_1,\mathcal{I}_2)$. Using \eqnref{eq:wPdyn} and \eqnref{eq:wP}, the Pauli weight is given by
\eq{w_{\scE_\sigma}(P)=\bigg(\prod_{i=1,2}(\bra{0}+\bra{1}/3)\bigg)\hat{T}\ket{w_{P,0}}.}
With the general form of the transfer matrix $\hat{T}$ in \eqnref{eq:gen form}, the shadow norm of a $k$-qubit Pauli operator, say $Z^{\otimes k}$ (with $k=0,1,2$ in a totally two-qubit system), is given by
\eq{
\Vert Z^{\otimes k}\Vert_{\scE_\sigma}^2=\frac{1}{w_{\scE_\sigma}(Z^{\otimes k})}=\left\{\begin{array}{ll}
1 &k=0,\\
\frac{9}{1+2(\mathcal{I}_1+\mathcal{I}_2)}&k=1,\\
\frac{27}{7-4(\mathcal{I}_1+\mathcal{I}_2)}&k=2,\\
\end{array}\right.}
which only depends on $\mathcal{I}_1+\mathcal{I}_2$ as shown in \figref{fig: gate}(b). 

One can see that the random Clifford  gates (at the ensemble average level) do not achieve the limit of minimizing the shadow norm for two-qubit Pauli observables in a two-qubit classical shadow tomography setup. There is still a little room for improvement by considering random unitary gates near the boundary of $\mathcal{I}_1+\mathcal{I}_2=1/3$, such as locally scrambled (i.e., Pauli twirled versions of) iSWAP gates. Local quantum circuits built from locally scrambled iSWAP gates are know to achieve the fastest scrambling speed \cite{2024arXiv241104898K}, making it advantages for reducing sample complexity in classical shadow tomography \cite{2024arXiv241201850W}. Given that the iSWAP gate is an important example of dual-unitary gates, it motivates us to investigate generic dual-unitary circuits more systematically in understanding its potential advantages in classical shadow tomography.

\subsection{Dual-unitary case}
\label{subsec:dual-unitary-transfer-matrix}

How does dual-unitarity affect or constrain this transfer matrix further? One may directly calculate $\mathcal{I}_1,\mathcal{I}_2$ by averaging over the choice of local scramblers in the parametrization of dual-unitary qubit gates. This gives the transfer matrix for dual-unitary gates in terms of $J$. Alternatively, we can ascertain the form of the transfer matrix directly via spatial unitarity. By the spatial unitarity condition, the operator entanglement $S^{(2)}_{|U\rangle\!\rangle}(AA')=2\log(2)$ is maximal. Therefore, $\mathcal{I}_1=0$. This is because dual-unitary gates maximize information scrambling. Given the understanding of the feasible parameter domain of $(\mathcal{I}_1,\mathcal{I}_2)$ in \eqnref{eq:I1I2bounds} (or see \figref{fig: gate}(a)), the dual unitary condition $\mathcal{I}_1=0$ has already restricted us to vertical boundary of the feasible domain with $1/3\leq \mathcal{I}_2\leq 1$.

In the following, we provide an independent explaination of the bound $1/3\leq \mathcal{I}_2\leq 1$ for dual-unitary gates without evoking the result cited from \refcite{You2018E1803.10425}. First of all, it's easy to check that in the dual-unitary case,
\begin{multline}\label{eq:enteqs1}
    S^{(2)}_{|U\rangle\!\rangle}(AB)=S^{(2)}_{|U\rangle\!\rangle}(A'B')\\=S^{(2)}_{|U\rangle\!\rangle}(AA')=S^{(2)}_{|U\rangle\!\rangle}(BB')=2\log(2),
\end{multline}
\begin{multline}\label{eq:enteqs2}
    S^{(2)}_{|U\rangle\!\rangle}(A)=S^{(2)}_{|U\rangle\!\rangle}(B)\\=S^{(2)}_{|U\rangle\!\rangle}(A')=S^{(2)}_{|U\rangle\!\rangle}(B')=\log(2).
\end{multline}
To determine the limits on $\mathcal{I}_2$, we need to bound $S^{(2)}_{|U\rangle\!\rangle}(AB')$. We can show that $S^{(2)}_{|U\rangle\!\rangle}(AB')\leq \log(2)$, by using the following inequality on four qubit pure states,
\begin{equation}
    I^{(2)}(A:A')+I^{(2)}(A:B')+I^{(2)}(A:B) \geq S^{(2)}(A),
\end{equation}
which says that the sum of the mutual information that $A$ shares with each other qubit is greater than its entropy. We can prove this equation through a known bound on the tripartite mutual information \cite{ibound}. Plugging in the definition for the mutual information on the left-hand side, 
\begin{align}
    &I^{(2)}(A:A')+I^{(2)}(A:B')+I^{(2)}(A:B) \\
    &=2S^{(2)}(A)+I(A:A':B)\\
    &\geq S^{(2)}(A),
\end{align}
where in the second line we plugged in the definition for the tripartite mutual information and in the last line we used that $I(A:A':B)\geq-I(A:A'|B)\geq -S^{(2)}(A)$.
Using \eqnref{eq:enteqs1}, \eqnref{eq:enteqs2}, we can see that (for dual-unitaries),
\begin{equation}
    0\leq S^{(2)}_{|U\rangle\!\rangle}(AB')\leq \log(2),
\end{equation}
which we can view as a version of entanglement monogamy. 
It follows that in the dual-unitary case, $1\geq \mathcal{I}_2\geq 1/3$. 

Given $\mathcal{I}_1=0$ and $1/3\leq \mathcal{I}_2\leq 1$ for dual-unitary gates, we can define the transfer matrix in terms of $\alpha:=1-\mathcal{I}_2$, where $0\leq \alpha\leq2/3$ is a parameter which characterizes the scrambling power of the dual-unitary gate. The general form of the transfer matrix in \eqnref{eq:gen form} reduces to the following for random two-qubit dual-unitary ensembles
\begin{equation}\label{eq:dutm}
    \hat{T}^{DU}_{b,b'}:=
    \begin{pmatrix}
     1 & 0 & 0 & 0 \\
     0 & 0 &  1-\alpha & \alpha/3 \\
     0 & 1-\alpha & 0 & \alpha/3 \\
     0 & \alpha & \alpha & 1-2\alpha/3
    \end{pmatrix}_{b,b'}.
\end{equation}
Alternatively, by direct calculation, we can show that
\begin{equation}
 \alpha=\frac{2}{3}\cos(2J)^2.   
\end{equation}
Therefore, at the swap point $J=\pi/4$, $\alpha$ vanishes. This is consistent with interpreting $\alpha$ as characterizing diffusive spreading of information in the circuit, since at this value of $J$ the circuit simply shifts the operator support around and does not actually diffuse the weight throughout the system. As a point of comparison, we also give the transfer matrix corresponding to random, two-qubit Clifford gates,

\begin{equation}\label{eq:cltm}
    \hat{T}^{Cl}_{b,b'}:=
    \begin{pmatrix}
     1 & 0 & 0 & 0 \\
     0 & 1/5 & 1/5 & 1/5 \\
     0 & 1/5 & 1/5 & 1/5 \\
     0 & 3/5 & 3/5 & 3/5
    \end{pmatrix}_{b,b'}.
\end{equation}

\subsection{Pauli Weight Dynamics}
\label{subsec:pauli-weight-dynamics}

\begin{figure}[H]
    \centering
    \includegraphics[width=0.8\columnwidth]{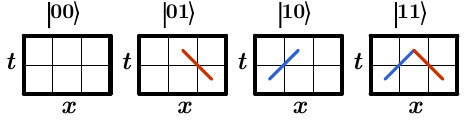}
    \caption{The basis states on two adjacent sites. For parity even sites $x+t\cong 0 \mod 2$ which support right movers, the basis states $\ket{0}$ and $\ket{1}$ are represented by the lack of a line and an upward slanting blue line, respectively. For parity odd sites $x+t\cong 1 \mod 2$ which support left movers, the basis states $\ket{0}$ and $\ket{1}$ are represented by the lack of a line and a downward slanting red line, respectively. }
    \label{fig:pw-basis}
\end{figure}

In this section, we hope to provide some intuition for the Pauli weight dynamics implied by the dual-unitary transfer matrix, \eqnref{eq:dutm}. Looking at the total weight one subspace, spanned by $b=10$ and $b=01$, we see that dual-unitarity implies that the diagonal elements vanish, and that the off-diagonal elements are always positive and greater than or equal to 1/3. The absence of the diagonal elements implies an interpretation of the dynamics in terms of left and right movers. We may organize the input legs of the transfer matrix (columns of \eqnref{eq:dutm}) so that the left (right) input bit indicates the presence of right (left) mover, and likewise we may organize the output legs of the transfer matrix so that the right (left) output leg corresponds to an outgoing right (left) mover i.e.
\begin{equation}
    \text{input: }
    \begin{cases}
    \emptyset & b'=00  \\
    \ell & b'=01  \\
    r & b'=10  \\
    \ell\cdot r & b'=11 
    \end{cases} \quad \text{output: }
    \begin{cases}
    \emptyset & b=00  \\
    r & b=01  \\
    \ell & b=10  \\
    \ell \cdot r & b=11 
    \end{cases}.
\end{equation}

We will see the value of this notational change when writing out the transfer matrix as a Markov process. We can visualize the basis states as follows: for a parity even (odd) site i.e. $x+t\cong 0\mod 2$ (1), the absence of a right-mover (left-mover) is corresponds to $\ket{0}$; the presence of a right-mover (left-mover) corresponds to $\ket{1}$ and is denoted by an upward slanting (downward slanting) blue (red) line. The basis states on two adjacent sites are depicted in \figref{fig:pw-basis}.

We see that dual-unitary dynamics preserve single left/right movers and they can only be absorbed in the presence of the opposite mover. Furthermore, we see that there are distinct processes by which the left and right movers can interact: they may branch, fuse, or collide, but there is no process by which a single left-mover and turn into a single-right mover. The rules are summarized in \tabref{tab:feynman_rules}. This is crucially different from the brick-wall Clifford case (with transfer matrix \eqnref{eq:cltm}), where the notion of a left or right mover is more fuzzy, since left movers can turn into right movers at one step. The specific locations of the left-right movers depending on the parity of the lattice position is specific to the staggered brick-wall set up of the circuit (see \figref{fig:cst}).

\begin{table*}[htbp]
\label{tab:rules}
    \centering
    \begin{tabular}{||m{15em} | m{15em} | m{15em}||}
        \hline
        \textbf{Process} & \textbf{Diagram(s)} & \textbf{Probability} \\
        \hline
        Vacuum conservation: The vacuum (no left or right movers incoming) is left invariant by the transfer matrix. & 
        \includegraphics[width=0.15\columnwidth]{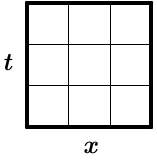} & $p(\emptyset\rightarrow\emptyset)=1$\\ \hline
        Propagation: A left-mover (right-mover) must continue to propagate as a left-mover (right-mover) in the absence of other movers. & \includegraphics[width=0.15\columnwidth]{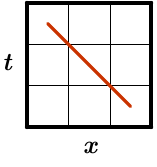} \includegraphics[width=0.15\columnwidth]{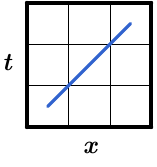} & 
        $p(\ell\rightarrow \ell)=p(r\rightarrow r)=1-\alpha$
        \\
        \hline
        Branching: A left-mover (right-mover) can split off a right-mover (left-mover) and continue to propagate. & \includegraphics[width=0.15\columnwidth]{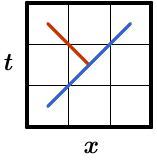} \includegraphics[width=0.15\columnwidth]{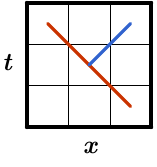} & 
        $p(\ell\rightarrow \ell \cdot r)=p(r\rightarrow \ell \cdot r)=\alpha$
        \\
        \hline
        Fusion: If a left-mover and right-mover meet, they can fuse into a single left or right mover.  &  \includegraphics[width=0.15\columnwidth]{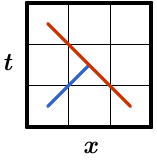} \includegraphics[width=0.15\columnwidth]{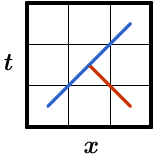} & 
        $p(\ell \cdot r\rightarrow \ell)=p(\ell \cdot r\rightarrow r)=\alpha/3$
        \\
        \hline
        Collision: If a left-mover and right-mover meet, they can go their separate ways.  & \includegraphics[width=0.15\columnwidth]{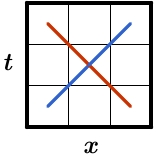} & 
        $p(\ell \cdot r\rightarrow \ell\cdot r)=1-2\alpha/3$
        \\ 
        \hline
    \end{tabular}
    \caption{The different processes allowed by the dual-unitary transfer matrix \eqnref{eq:dutm} and their corresponding probabilities. Here, the left and right legs are colored differently to distinguish the types of particles but both correspond to a $\ket{1}$ state on their corresponding leg of the transfer matrix. }
    \label{tab:feynman_rules}
\end{table*}

When $\alpha=0$, the branching and fusing is disallowed and the left or right movers propagate freely without being disturbed. In open boundary conditions, they may change parity by bouncing off the walls, whereas in periodic boundary conditions, they simply come back around. When $\alpha>0$, there is some probability that these particles can interact via fusing or branching. Note that the particle symmetry is explicitly broken when $\alpha>0$, and we see that non-zero $\alpha$ will condense to some equilibrium density of particles.  

The notion of left/right movers is also useful because it helps us understand the light-cone structure that arises in the case of dual-unitary Pauli weight dynamics. If a left mover is able to avoid being absorbed by any right movers to the left of it, or it is emitted by the left-most right-mover, it will propagate indefinitely until it reaches the system boundaries. The motion of this particle defines the left light-cone boundary. This lightcone structure is intimately related to the fact that dual-unitary brick-wall circuits are an example of quantum cellular automata, see for e.g., the excellent review \cite{Farrelly2020reviewofquantum}.

\subsection{Occupation distribution}
\label{subsec:occupation-distribution}

\begin{figure}[H]
    \centering
    \includegraphics[width=0.49\columnwidth]{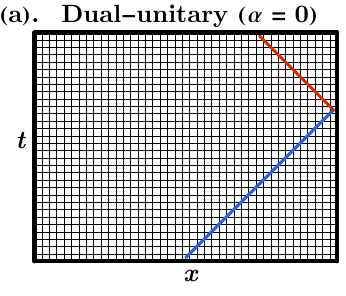} 
    \includegraphics[width=0.49\columnwidth]{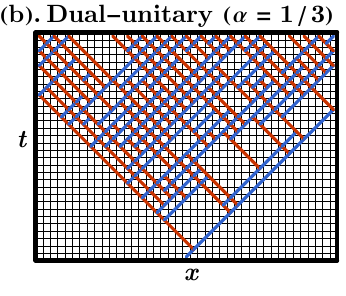}\\
    \includegraphics[width=0.49\columnwidth]{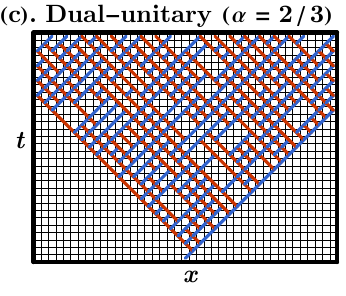}
    \includegraphics[width=0.49\columnwidth]{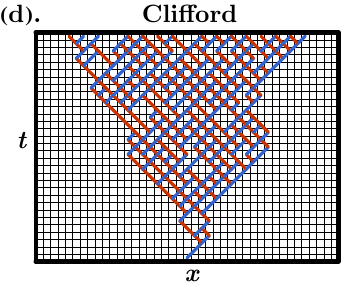} 
    \caption{(a-c). A typical Pauli weight distribution evolved by the dual-unitary transfer matrix \eqnref{eq:dutm} for $\alpha=0,1/3,$ and $2/3$, respectively. When $\alpha$ is small, the probability of branching is small as well, so the initial right-mover basically moves freely. As we increase $\alpha$, we increase the probability of branching so we see an earlier emergence of a left light-cone, and we see more uniform particle density inside the light-cone. (d). Typical Pauli weight distribution for the Clifford case. There is no light-cone in this case.}
    \label{fig:samples}
\end{figure}

In this section, we define simple order parameters of the Pauli weight as a distribution to analyze the average occupation over different sub-regions of sites. In addition, we study Monte Carlo simulations to gain some intuition about the Pauli weight dynamics in dual-unitary circuits. We can view the Pauli weight dynamics generated by the transfer matrix in \eqnref{eq:dutm} as a Markov process that defines how a configuration evolves with time. For example, if the configuration on two adjacent sites consists of an incoming left-mover and right-mover, then the outgoing configuration will consist of a single left-mover with probability $\alpha/3$, a single right-mover with probability $\alpha/3$, and both left-mover and right-mover with probability $1-2\alpha/3$. Using the rules implied by the transfer matrix, we can generate Pauli weight configurations prior to the final measurement layer. These configurations consist of a $n \times t$ grid, where $n$ is the number of qubits in the Pauli weight Hilbert space and $t$ is the overall circuit depth. Each point in the grid can take a value of one or zero, depending on if a given site contains a particle or not. These configurations can be generated using Monte Carlo techniques, and some typical configurations are depicted in \figref{fig:samples}. Each configuration begins with a single Pauli operator localized at the center of the system $x=n/2$. In these figures, we color the even/odd parity sites differently to more easily differentiate sites which host left-movers from those that host right-movers.

\begin{figure}[!h]
    \centering
    \includegraphics[width=0.49\columnwidth]{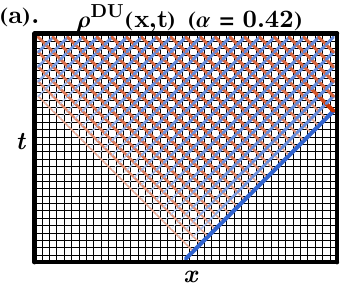}
    \includegraphics[width=0.49\columnwidth]{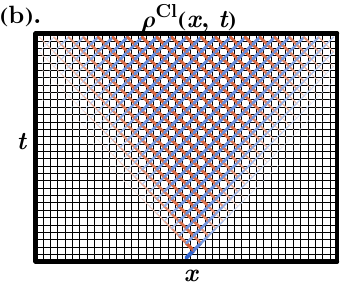}
    \caption{The occupation $\rho(x,t)$ displayed on a grid of size $n=40,t=30$, starting from an initial configuration $\rho(x,0)=\delta_{x,n/2}$. Even and odd sites are distinguished by blue and red colors, respectively. Thicker lines indicate a larger occupation at $(x,t)$. (a). The average occupation for $\alpha=0.42$. Since $\alpha$ is non-negligible, branching and fusing leads to a large weight inside the light-cone. The shape of the light-cone is visible because the probability of emitted the first left-mover is $\alpha$. (b). The occupation in the Clifford dynamics case. Because of the instability of chiral excitations, the light-cone boundaries are less well-defined and the average occupation goes to zero smoothly as we move towards the system boundaries.}
    \label{fig:avg_wgt_k=1}
\end{figure}

There are some interesting features of the dual-unitary case that we would like to bring to the readers' attention. First, we see that the dual-unitary case (\figref{fig:samples} a-c) inherits a light-cone structure \cite{Farrelly2020reviewofquantum}. Because we only start with a single-particle at $t=0$, depending on the parity of the initial site, the injection of the this particle will create a forever-propagating excitation that cannot be absorbed by anything (until it hits the boundary). In \figref{fig:samples} a-c, since the particle is always injected on a parity even site, there is always a consistent right light-cone boundary. The first left-mover that branches from this initial particle injection will form a left light-cone boundary also defined by a forever-propagating excitation. When $\alpha$ is small, the probability of emitting this left-mover is small, and so the right-mover mostly propagates freely. When $\alpha$ is large, there is a higher probability of branching and fusing, which creates a condensate of of particles within the light-cone. 

This behavior is in contrast to \figref{fig:samples} d, which is generated by the random Clifford transfer matrix. According to this transfer matrix, there is a non-negligible probability that a single left-mover may be turned into a single right-mover, and vice versa. This means that the Pauli-weight spreads ``diffusively'' rather than ballistically as in the dual-unitary case. Therefore, we don't see any light-cone in individual samples.

To refine our analysis, recall that the Pauli weight defines a probability distribution over all possible sub-regions of $\Omega=\{1\dots n\}$, or equivalently, all possible $n$-bit strings (see \eqnref{eq:bit-region-mapping}. This is because the transfer matrix which evolves the Pauli weight vector defined in \secref{sec:setup} is a probability-conserving process. Therefore, we have that the weights are positive i.e. $w_{P,t}(A) := \braket{b}{w_{P,t}}>0$, where $b$ is related to $A$ by \eqnref{eq:bit-region-mapping}, and also that 
\begin{equation}
    \sum_{A\subset \Omega}w_{P,t}(A)=1.
\end{equation}
Let us define the \textit{average occupation} of the Pauli weight on region $A\subset \Omega$ at time $t$, $\rho(A,t)$, as the probability that particles reside on all sites $i\in A$ simultaneously,
\begin{equation}
    \rho(A,t):=\sum_{B\supset A}w_{P,t}(B).
\end{equation}
This equation says that the average occupation is given by summing over the possible occupations on sites $i\notin A$, and fixing a particle for all sites $i\in A$. Likewise, this expression may be inverted to give the probably $w_{P,t}(A)$ via
\begin{equation}\label{eq:occ-to-pw}
    w_{P,t}(A)=\sum_{B\supset A}(-1)^{|A|-|B|}\rho(B,t).
\end{equation}
The main object of study will be the average occupation on site $x$ at time $t$, $\rho(x,t):=\rho(\{x\},t)$. This can be evaluated numerically by taking the average of all the grid configurations generated by Monte Carlo sampling. In terms of $\ket{w_{P,t}}$, the Pauli weight state at time $t$, 
\begin{align}\label{avg_weight}
    \rho(x,t)&=\bigotimes_{i=1\dots n}\left( \bra{0}(1-\delta_{i,x}) + \bra{1} \right)  \ket{w_{P,t}},
\end{align}
which says that we sum over weight configurations on every site other than $x$. The average occupation in the dual-unitary cases differs sharply from the Clifford case because of the emergent light-cone. In \figref{fig:avg_wgt_k=1} a, we see the average occupation in the dual-unitary case. In contrast, looking at \figref{fig:avg_wgt_k=1} b for the average occupation in the Clifford case, we see that that the weight smoothly goes to zero as we move towards the system boundaries.

\begin{figure}[!t]
    \centering
    \includegraphics[width=0.49\columnwidth]{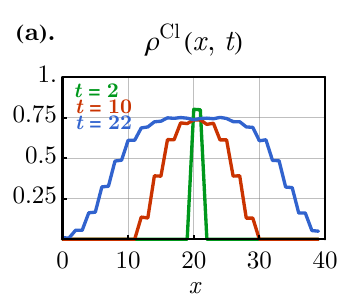} 
    \includegraphics[width=0.49\columnwidth]{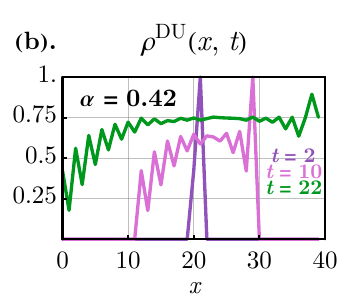} 
    \includegraphics[width=0.49\columnwidth]{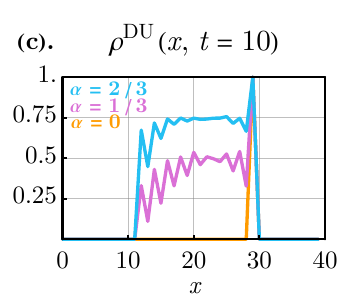} 
    \caption{Time slices for the average occupation $\rho(x,t)$ for system parameters $n=40,t=30$, starting from an initial configuration $\rho(x,0)=\delta_{x,n/2}$. (a). The average occupation $\rho(x,t)$ in the Clifford case at fixed time slices $t=2,10,22$. The average occupation spreads diffusively in both directions with no overall bias. The weight smoothly goes to zero at both boundaries. Even-odd sites interpolate smoothly between each-other. The average occupation peaks at $3/4$. (b). The average occupation $\rho(x,t)$ for $\alpha=0.42$ at fixed time slices $t=2,10,22$. In this case, the weight has a sharp spike at the location of the right light-cone boundary: $\rho(n/2+t,t)=1$. Inside the light-cone, there is a sharper dependence on the site parity and greater uniformity of the average occupation. For example, for $t=22$, the weight is mostly uniform spatially at the equilibrium value of $3/4$. (c). We investigate a fixed time slice at $t=10$ for various values of $\alpha$. We find that as we increase $\alpha$, the height of the average occupation increases and eventually saturates at $3/4$.}
    \label{fig:slices_k=1}
\end{figure}

In \figref{fig:avg_wgt_k=1} a, the average occupation $\rho(x,t)$ starting from a single-particle configuration in the dual-unitary case has a clear right boundary due to the initial injection of the right-mover. Because of this, the spatial structure of the average occupation at fixed time-slices differs markedly from the Clifford case. In particular, because the initial injected right mover cannot be absorbed by a left-mover, the weight on the right light-cone is always one, whereas the weight goes to zero as we approach the left-boundary. This can be seen most acutely in \figref{fig:slices_k=1}. 

Looking at \figref{fig:slices_k=1} a, we see that in the brick-wall Clifford case, the average occupation spreads in an unbiased fashion over time, initially localized at near the initial Pauli support. As we evolve in time, the weight becomes more spread out spatially, with the bulk of the weight taking a roughly uniform value of 3/4. This is because deep Clifford circuits eventually thermalize and model global unitaries, which scramble Pauli operators uniformly. Out of all Pauli strings, 3/4 of them have support on site $x$, so the average occupation $\rho(x,t)$ converges is 3/4 inside the effective light-cone.

Contrarily, \figref{fig:slices_k=1} b breaks the inversion symmetry in the occupation. Because the initial Pauli weight configuration consists of a single right-mover, this particle will propagate with velocity $v=1$. Furthermore, the location of the right mover always has an average occupation of one. Therefore, we see a spike at the location of the right light-cone boundary. As we move away from the location of this right-mover, the weight goes to zero. In the dual-unitary case, we see there is much greater fluctuation as we move from even to odd parity sites and vice versa. In \figref{fig:slices_k=1} c, we see that that $\alpha$ characterizes how quickly the inside of the light-cone thermalizes. For a fixed time-slice, as we increase $\alpha$, we increase the average value of the weight inside the light-cone. At $t=10$, we see that when $\alpha=2/3$, the weight is uniformly 3/4. However, as we decrease $\alpha$, the weight at the same value of $t$ is smaller.

\begin{figure}[!t]
    \centering
    \includegraphics[width=0.55\columnwidth]{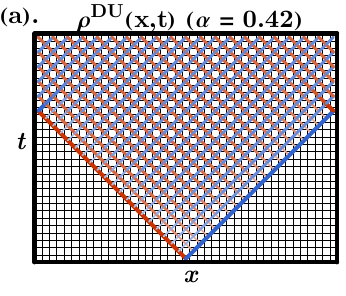}
    \includegraphics[width=0.49\columnwidth]{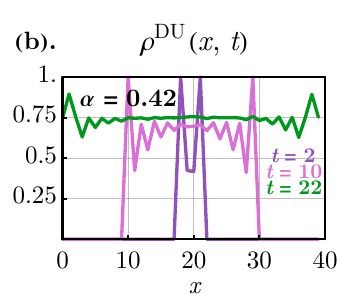}
    \includegraphics[width=0.49\columnwidth]{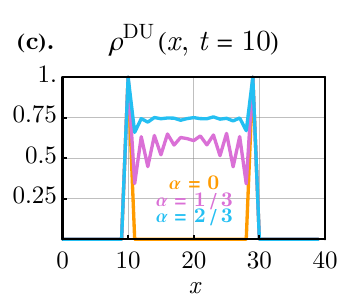}
    \caption{Time slices for the average occupation $\rho(x,t)$ for system parameters $n=40,t=30$, starting from an initial configuration $\rho(x,0)=\delta_{x,n/2-1} + \delta_{x,n/2} $. (a). Visual representation for the space-time dependence of the average Pauli weight. Inside the light-cone, the weight appears mostly uniform, whereas on the boundaries it takes value one. (b). The average occupation $\rho(x,t)$ for $\alpha=0.42$ at fixed time slices $t=2,10,22$. There are spikes at the location of the initial left/right movers. There is an initial drop moving inside the light-cone, which then gradually becomes larger as we approach the center of the light-cone. (c). As we increase $\alpha$ for fixed time slice $t=10$, the weight inside the light-cone increases towards the saturation value of 3/4. }
    \label{fig:slices_k=2}
\end{figure}

We can restore the inversion symmetry by injecting a left-mover immediately to the left of the initial right mover. This is seen in \figref{fig:slices_k=2}. Now there is a left light-cone boundary defined by the initial left-mover. At both the left and right boundary, the weight is always one. To understand the rich behavior of the Pauli weight within the light-cone, we will develop a mean-field analysis of the average weight dynamics.

\subsection{Mean Field Approximation of Pauli Weight Dynamics}
\label{subsec:mean-field}

According to \eqnref{avg_weight}, the time evolution of $\rho(x,t)$ is determined through the evolution of the many-body Pauli weight state $\ket{w_{P,t}}$, which happens through the application of a brick-wall arrangement of transfer matrices \eqnref{eq:dutm}  (depicted graphically in \figref{fig:cst}b). Since the exact evolution of the many-body state $\ket{w_{P,t}}$ is complicated, this does not give us a simple description of the evolution of $\rho(x,t)$. In fact, there is no exact, closed-form equation for the evolution for $\rho(x,t)$ because in general the average occupation at a single site will depend on ``higher-order'' contributions from various other average occupation configurations on multiple sites. We instead look for a simplified approach for the evolution of $\rho(x,t)$ inspired by mean-field theory.

One approach is to ignore the correlations between individual sites. Since it is known that the entanglement features has an efficient low-bond dimension description, the same can be expected of the Pauli weight which is entanglement features by a local basis transformation \cite{akhtar2023measurementinduced}. The reason these many-body vectors have low entanglement is because of their positive sign structure \cite{Grover2015E,Akhtar2023S2209.02093}. We can specify the $k$-th order approximation, $k\leq n$, by saying that all $l$-particle occupations, $l\leq k$, factorize: 
\begin{equation}\label{eq:main-approx}
    \rho(A)\approx\prod_{x\in A}\rho(x) \quad \forall A \text{ s.t. } |A|\leq k.
\end{equation}
To derive an equation of state of the single site occupancy $\rho(x,t)$, we only need to use the approximation to second order i.e. $k=2$.  

The idea is to derive an equation of state for $\rho(x,t)$ from the transfer matrix in \eqnref{eq:dutm}. Let $\rho^L_{\text{out}}$ refer to the left, out-going leg of a two-local transfer matrix like in \figref{fig:cst} (b). The weight $\rho^L_{\text{out}}$ depends on the incoming weight. We can approximate the probability that the incoming weight configuration include only one left mover by $\rho^L_{\text{in}}(1-\rho^R_{\text{in}})$; the probability that the incoming weight configuration includes only a right mover by $\rho^R_{\text{in}}(1-\rho^L_{\text{in}})$; and the probability that the incoming weight configuration consists of both a left and right mover by $\rho^L_{\text{in}}\rho^R_{\text{in}}$. If there is exactly one incoming left-mover, then there must be an outgoing left-mover; if there is only one incoming right-mover, then there is a out-going left-mover with probability $\alpha$; if both a left-mover and right-mover are incoming, then with probability $1-\alpha/3$ there is an out-going left-mover. Putting it all together, we get an equation for $\rho^L$ and an equivalent equation for $\rho^R$ using the same logic:
\begin{equation}\label{eq:mfa}
    \begin{aligned}
    &\rho^L_{\text{out}}=\rho^L_{\text{in}}+\alpha( \rho^R_{\text{in}}-\frac{4}{3}\rho^L_{\text{in}}\rho^R_{\text{in}}),\\
    &\rho^R_{\text{out}}=\rho^R_{\text{in}}+\alpha( \rho^L_{\text{in}}-\frac{4}{3}\rho^L_{\text{in}}\rho^R_{\text{in}}).
    \end{aligned}
\end{equation}

\begin{figure}[H]
    \centering
    \includegraphics[width=0.8\columnwidth]{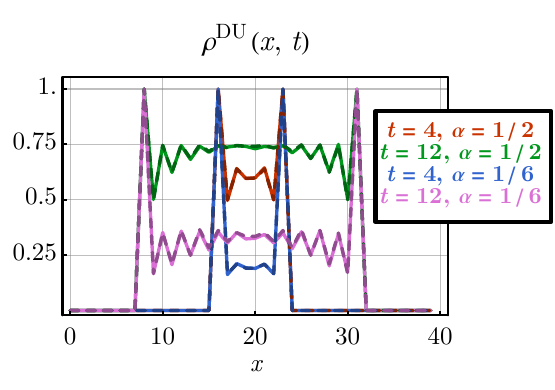}
    \caption{We demonstrate the accuracy of \eqnref{eq:mfa}, which follows from \eqnref{eq:main-approx} for $k=2$, in simulating the single-site occupation. We find practically no discernible difference for various choices of $\alpha,t,$ etc. In the figure above, the solid lines are generated using the Monte Carlo simulations of the Pauli weight, whereas the dashed lines are generated using \eqnref{eq:mfa}, for a system of size $n=40$, for $\alpha=1/6,1/2$ at times $t=4,12$. We find that for simulating the single-site occupation $\rho(x,t)$, there is no loss of accuracy in using the $k=2$ mean-field approximation.}
    \label{fig:mft-test}
\end{figure}

We find that simulating the above equation to update the average occupation $\rho(x,t)$ is practically indiscernible from the average occupation determined through Monte Carlo simulations. Therefore, the mean-field approximated dynamics for the average occupancy is an accurate description for the average occupation $\rho(x,t)$ for all choices of $\alpha$. This is demonstrated numerically in \figref{fig:mft-test}.

The two equations \eqnref{eq:mfa} can be written as one equation depending on the parity of the space-time location $x,t$. If $x+t\equiv 0\mod 2$ ($x+t\equiv 1\mod 2$), then the site is parity odd and hosts a left-mover. The average occupation at the next layer is given as a single recurrence relation: 
\begin{multline}\label{eq:mfa_xt}
    \rho(x,t+1)-\rho(x+v_{x,t},t)\\=\alpha\rho(x,t)\left(1-\frac{4}{3}\rho(x+v_{x,t},t)\right),
\end{multline}
where $v_{x,t}=(-1)^{x+t}.$ To obtain a qualitative understanding of the recurrence relation, we can reinterpret this as a differential equation by approximating the discrete differentials as continuous derivatives. If we then drop all second order or higher terms, we end up with a first order partial differential equation for $\rho$ that is indexed by the sign of the velocity,
\begin{equation}\label{eq:mfa_dirac}
    \partial_t\rho- v(\rho)\partial_x \rho = m(\rho) \rho,
\end{equation}
where we've introduced a mass function $m(\rho)$ and corresponding velocity $v(\rho)$ given by
\begin{align}
    m(\rho)&:=\alpha'(1-4\rho/3), \\  
    v(\rho)&:=f(x+t)(1-4\alpha'\rho / 3),
\end{align}
where $f$ is a space-time function that modulates the changing sign of velocity in the lattice. For example, one choice for $f$ could be $f(s)=\cos(\pi s)$, since this takes the correct value at the lattice points, but this choice is not unique and in principle should not affect the low-energy physics. We also replace $\alpha\rightarrow \alpha'$ to accommodate the fact that approximating a quadratic recurrence relation by a continuous differential equation can in general produce discrepancies. The mass function $m(\rho)$ decreases linearly with $\rho$, going from $m(0)=\alpha'$ to $m(1)=-\alpha'/3$, passing through a zero at the saturation value $\rho=3/4$ i.e. $m(3/4)=0$. Contrarily, the speed $|v(\rho)|$ decreases from $|v(0)|=|f(x+t)|$ to $|v(1)|=|f(x+t)(1-4\alpha'/3)|$, passing through $|f(x+t)(1-\alpha')|$ at the massless point $\rho=3/4$. 

We label the variables $m$ and $v$ in such a way to highlight the similarity to the Dirac equation. We see that $m(\rho)$ plays the role of the fermion mass. As the fermion becomes massive, it warps the space around it, which is accomplished through the velocity coefficient $v(\rho)$. If we consider a region of spacetime where $m(\rho)$ is slowly varying, such that we can treat $m(\rho)$ and $v(\rho)$ as constants, the mean-field equation matches one of the chiral components of the Dirac field.

\begin{figure}[H]
    \centering
    \includegraphics[width=0.6\columnwidth]{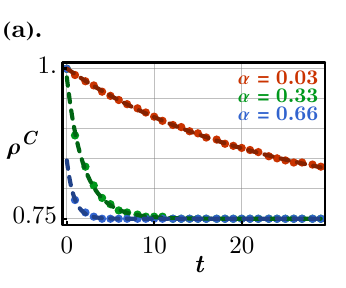}
    \includegraphics[width=0.6\columnwidth]{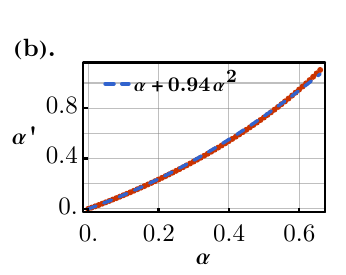}
    \caption{(a). We compare the homogeneous solution $\rho^C$ in the continuum \eqnref{eq:constcase} with the exact, numerical solution generated by \eqnref{eq:constcase2}. The exact, numerical values are represented by dots, whereas the plot of \eqnref{eq:constcase} for the corresponding value of $\alpha'$ is given by the dashed lines. We find exact agreement for appropriately chosen $\alpha'$ for all $\alpha$. (b). We compare $\alpha$ with $\alpha'$ as determined by fitting $\alpha'$ in \eqnref{eq:constcase} according to the exact numerics. We find that $\alpha'$ depends non-linearly on $\alpha$, deviating by a quadratic correction as $\alpha$ is becomes larger.}
    \label{fig:avaeff}
\end{figure}

The equation \eqnref{eq:mfa_dirac} is non-linear since the coefficients $v$ and $m$ themselves depend on the solution $\rho$. Let us try to solve \eqnref{eq:mfa_dirac} in various settings. First, consider the case where the spatial derivative $\partial_x \rho$ vanishes. This could happen because $\rho$ is homogeneous, or it could happen along a line of spatial-inversion symmetry. More commonly, this is also expected to hold within the bulk of the support of a contiguous operator, as can be observed in \figref{fig:mft-test}. This is because initially, at $t=0$, the occupation takes value one so there is no fluctuation within the bulk $\partial_x \rho=0$. As we increase the circuit depth $t$, the boundaries of the contiguous operator will propagate outwards with the light-velocity, and the brick-wall structure of the circuit will create smaller-scale fluctuations within the unit cell. However, over longer scales the velocity term in \eqnref{eq:mfa_dirac} will average out and only change $\partial_x \rho$ gradually. Label $\rho=\rho^C$ in the region where $\partial_x \rho=0$, then 
\begin{equation}\label{eq:center}
    \partial_t\rho^C=\alpha'\rho^C (1 -\frac{4}{3}\rho^C).
\end{equation}
Here, we've substituted in the exact expression for $m(\rho)$. This differential equation may be solved easily by substituting e.g. $u=1/\rho^C$. The result is 
\begin{equation}\label{eq:constcase}
    \rho^C(t)=\frac{3}{4}\big/(1+Ae^{-\alpha' t})
\end{equation}
where $A$ is a constant. Thus, the weight along a center line or homogeneous region tends exponentially towards the equilibrium value of $3/4$ with exponent $\alpha'$. We also see that  $\alpha'$ tunes the relaxation rate of the Pauli weight, with the system thermalizing more quickly when $\alpha'$ is larger.

We can compare \eqnref{eq:constcase} against $\rho$ determined by the exact recurrence relation when $\rho$ is spatially homogeneous,
\begin{equation}\label{eq:constcase2}
    \rho(t+1)=\rho(t)+\alpha\rho(t)(1-4\rho(t)/3),
\end{equation}
with initial boundary condition $\rho(0)=1$. We find that \eqnref{eq:constcase} captures the exact behavior of $\rho$ in the homogeneous case accurately for $t\geq1$. We find, however, that $\alpha'$ becomes a quadratic function of $\alpha$ (see \figref{fig:avaeff}). The quadratic corrections to $\alpha'$ arise because the continuum approximation, in which we are replacing the discrete differences with a first derivative, breaks down when $\alpha$ is large. However, the corrections become negligible when $\alpha$ is small so that $\alpha'\rightarrow\alpha$ when $\alpha\rightarrow 0$, as expected.

\subsection{Ramifications for the shadow norm and prediction}
\label{subsec:shadow-norm}

The average occupation $\rho(x,t)$ at the end of the circuit evolution can be used to approximate the shadow norm, which characterizes the sample complexity of the corresponding classical shadow scheme (see \figref{fig:cst}). In order to transform $\rho(x,t)$ into a shadow norm, we must first incorporate the effects of the final measurement layer on the average Pauli weight. A local measurement weights particles (holes) by 1/3 (1). Therefore, in terms of the final Pauli weight $w_{P,t}(A)$, the Pauli weight of the classical shadow protocol $w_{\mathcal{E}_\sigma}(P)$ is given as a weighted sum over all components. Equivalently, we may re-write this sum in terms of the final multi-region occupations $\rho(A,t)$ using \eqnref{eq:occ-to-pw}
\begin{equation}
    w_{\mathcal{E}_\sigma}(P)=\sum_{A\subset \Omega} (-2/3)^{|A|}\rho(A).
\end{equation}
where in the second equality, we've suppressed the time coordinate because it is assumed that we are considering only the final time slice before the application of the measurement gates. To simplify our analysis, we invoke the approximation \eqnref{eq:main-approx} at $n$-th order: $\rho(A)=\prod_{i\in A}\rho(x)$ for all non-empty regions $A$. This approximation ignores the correlations in the occupation numbers at different sites. While this is a reasonable approximation for the individual occupancies $\rho(A)$, because the Pauli weight vector carries little entanglement due to its limited sign structure \cite{Grover2015E,Akhtar2023S2209.02093,Akhtar2020M2006.08797}, making the substitution overestimates the Pauli weight. The discrepancy is because the shadow norm involves the sum of an exponential number of occupations, which can contribute in total to a small error in the thermodynamic limit. Performing the substitution gives the right hand side of the following equation:
\begin{equation}\label{eq:aw2pw}
    w_{\mathcal{E}_\sigma}(P) \approx \prod_{x=1\dots n}\left(1-2\rho(x)/3\right),
\end{equation}
where $\rho(x)$ is assumed to be the single-site occupancy after the final unitary layer. Strictly speaking, the right hand side of \eqnref{eq:aw2pw} \textit{upper bounds} the Pauli weight when the support of $P$ is large with respect to the system size. To understand why there is an upper bound, first rewrite the average occupation as an expectation value over indicator variables $x_i$. The idea is that $x_i=0$ or $1$ depending on if there is a particle on site $i$ or not. The probability of a particular configuration of $\{x_i|i=1\dots n\}$ is given by the corresponding Pauli weight:
\begin{equation}
    p(x_1,\dots,x_n)=w_{P,t}(\{i\,| \,x_i=1\})
\end{equation}
Then, both sides of \eqnref{eq:aw2pw} may be rewritten as an expectation value over the binary indicator variables $x_i$ distributed according to $p(x_1\dots x_n)$. The bound says that these indicator variables are correlated, namely
\begin{equation}
    \mathbb{E}[\prod_i (1/3)^{x_i}]\leq \prod_i\mathbb{E}[ (1/3)^{x_i}].
\end{equation}
This bound can be seen to follow in the case of finite correlation length in the indicator variables. In this case, the sum of the indicator variables in the exponent on the left hand side will generally be sharply peaked around the mean $\sim n \rho(x)$, assuming that the correlation length is finite. Hence, the left hand side will generally scale exponentially with a base of $\sim (1/3)^{\rho(x)}$, whereas the right hand side will scale exponentially with a larger base of $1-2/3\,\rho(x)$. See the appendix \secref{app} for a more rigorous argument in the case where $P$ has full support and an analysis of the accuracy of the approximation in characterizing the scaling of the Pauli weight.

The shadow norm is given by the inverse Pauli weight. Using the same trick, we get an approximate lower-bound on the shadow norm:
\begin{equation}\label{eq:aw2sn}
    \| P \|^2_{\mathcal{E}_\sigma} 
    \approx \prod_{x=1\dots n} \beta(x), \,\, \beta(x)=\left(1-2\rho(x)/3\right)^{-1},
\end{equation}
with the scaling exponent of the shadow norm given by $\beta(x)$. Although the approximation technically gives a lower bound on the shadow norm, we find that it is still useful in characterizing the qualitative behavior of the Pauli weight in the remainder of this section. Furthermore, we find that the analytical form implied by \eqnref{eq:aw2sn} fits our numerical data with a renormalized $\alpha$, for small $\alpha$. 

The essential physics of the Pauli weight dynamics can be captured by two competing effects: the ballistic spreading of non-trivial Pauli weight, and the saturation of the average occupation towards the equilibrium value of 3/4. When $t=0$, the shadow norm scales as $\|P\|^2= 3^k$ where $k$ is the size of the support of $P$. This is expected for any locally-scrambled ensemble at zero depth. For deeper circuits $t>0$, the initial support will spread out, while the bulk weight saturates. The specific way in which this happens depends on the type of unitary evolution. Nonetheless, since $\beta(x)\rightarrow 2$ as $\rho(x)\rightarrow 3/4$, we see that thermalizing locally-scrambled evolution always tends towards Clifford measurements i.e. $\|P\|^2 \sim 2^n$ for large circuit depth $t$ irrespective of the initial support of $P\neq \id$. While the relaxation of the base $\beta(x)$ decreases the shadow norm, the spreading of the average weight in general increases the shadow norm. This is because only the sites with $\rho(x)>0$ contribute to the shadow norm. Therefore, when $t=0$, the weight is $3^k$, but as the circuit evolves $t>0$, the shadow norm behaves as 
\begin{equation}
    \|P\|^2\sim (\beta_{\text{eff}})^{k_{\text{eff}}}
\end{equation}
where $2 \leq \beta_{\text{eff}}\leq 3$ and $k\leq k_{\text{eff}}\leq k+2t$ describe the effective base and size of the evolved operator, respectively. The initial decrease in the base for Clifford gates is why shallow depth circuits can provide an advantage over Pauli measurements. For brick-wall Clifford circuits, the overall shadow norm decreases until the optimal depth $t^{*}\sim \log k$, but increasing the circuit depth further actually increases the shadow norm because of $k_{\text{eff}}$ becomes too large \cite{Ippoliti2023O2212.11963}.

\begin{figure}[!t]
    \centering
    \includegraphics[width=0.7\columnwidth]{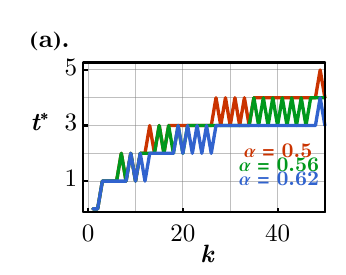}
    \includegraphics[width=0.7\columnwidth]{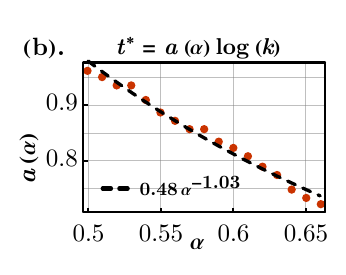}
    \caption{(a). We numerically calculate the optimal circuit depth $t^{*}$ for contiguous Pauli strings of length $k$ for different values of $\alpha$ in a system of $n=100$ total qubits. Larger $\alpha$ thermalizes the system more quickly, thereby lowering $t^{*}$. (b). We determine the fitting parameter $a(\alpha)$ to the functional form $t^{*}=a(\alpha)\log(k)$ using a simple least-squares method. We then fit $a(\alpha)$ as a function of $\alpha$ to the functional form $a(\alpha)=c \alpha^b$, and find that $b\approx -1$ as predicted by the mean-field theory. }
    \label{fig:du_opt_depth}
\end{figure}

Similar behavior can be observed in dual-unitary circuits as well. In general, if we consider the variation of $\log \|P\|^2$ with respect to the circuit depth $t$, we see that there are two contributions: 
\begin{equation}
    \delta \log \|P\|^2 = \delta k_{\text{eff}} \log\beta_{\text{eff}} + k_{\text{eff}}\delta\beta_{\text{eff}}/\beta_{\text{eff}}.
\end{equation}

The first term $ \delta k_{\text{eff}} \log\beta_{\text{eff}}$ is a positive contribution which scales like the boundary of the support of $P$. For contiguous Pauli strings $P$, this is a constant contribution that does not scale with the size of the operator. The second term $k_{\text{eff}}\delta\beta_{\text{eff}}/\beta_{\text{eff}}$ is a negative contribution (since $\delta\beta_{\text{eff}}$ is negative) that scales like the size of the support of $P$. Initially, the drop in $\beta_{\text{eff}}$ is large and the second term dominates, meaning its favorable to increase the circuit depth. However, as $\beta$ approaches two, $\delta\beta_{\text{eff}}/\beta_{\text{eff}}$ becomes small in magnitude and the first term dominates, meaning that its disadvantageous to increase the circuit depth further. These terms are comparable when $|\beta_{\text{eff}}/\delta \beta_{\text{eff}}|\sim k_{\text{eff}}$. 

According to the mean field approximation \eqnref{eq:aw2sn}, $|\beta/\delta\beta|\sim 3/(2\beta \delta \rho)$. From \eqnref{eq:mfa_xt}, we see that the weight will generally decrease exponentially towards saturation value, with the constant $\alpha$ setting an overall scale for the difference in the weight as we move deeper diagonally in the circuit. Simple scaling analysis suggests that
\begin{equation}
    t^{*}\sim \log(k) / \alpha.
\end{equation}
This scaling dependence on $k$ and $\alpha$ is confirmed numerically. We do this by simulating the shadow norm for the dual-unitary circuit using a large bond dimension MPS, thereby eliminating any truncation error, in a large $n$ system. We find that the optimal circuit depth scales logarithmically in the operator size and is inversely proportional to $\alpha$, as predicted by the mean-field analysis.

\begin{figure*}[t]
    \centering
    \includegraphics[width=0.65\columnwidth]{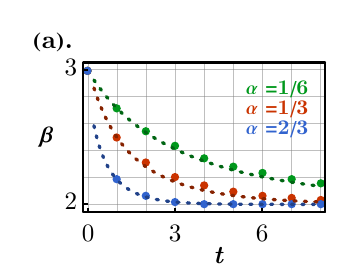}
    \includegraphics[width=0.65\columnwidth]{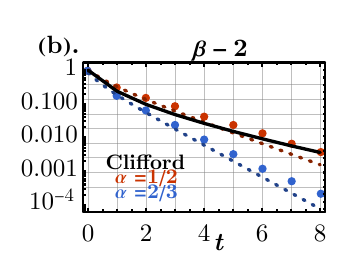}
    \includegraphics[width=0.65\columnwidth]{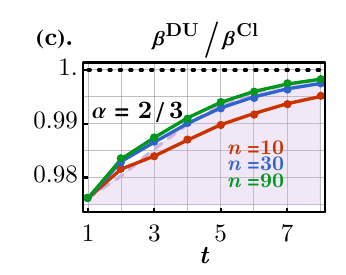}
    \caption{We simulate the scaling of the shadow norm on a $n$-qubit Pauli operator, where $n=100$ in (a-b), using MPS simulations of the Pauli weight for a bond dimension of $D=200$, at finite circuit depth $t$. We numerically define $\beta=(\|P\|^2)^{1/n}$ as how the shadow norm scales with system size. This is represented by the dots in the above figure. The predictions from the mean-field theory \eqnref{eq:beta-mft-cont} are represented by the dotted lines. (a). We compare the scaling of the shadow norm $\beta$ for a fully extensive operator with the predictions from the mean-field theory for various values of $\alpha$. We find strong agreement between the two. (b). We compare the approach to the Clifford measurement limit of dual-unitary circuits to brick-wall Clifford circuits represented by the black line. We find that dual-unitary circuits out can perform brick-wall Clifford circuits for full support operators, approaching the Clifford measurement limit more quickly. For $\alpha=2/3$, there is an advantage at all circuit depths. (c). We compare brick-wall dual-unitary circuits with brick-wall Clifford gates at finite size for operators with extensive support. The mean-field approximation, which lower bounds the shadow norm in the thermodynamic limit, is represented by the boundary of the shaded region. We find that for small systems, the advantage of using dual-unitary gates over Clifford gates \textit{increases}. This is because the boundary effects in the dual-unitary case are favorable for dissipating excitations in the occupation, thereby lowering the shadow norm. We also find that the exact numerics converge to the mean-field approximation result above $n\approx 30$, whereas below $n\approx 30$ the mean-field approximation sets a reliable upper bound to the shadow norm.}
    \label{fig:du_sat_rate}
\end{figure*}

In general, when $k_{\text{eff}}$ is smaller than the system size $n$, it is better to use brick-wall Clifford circuits for prediction. This is because of the rapid growth of $k_{\text{eff}}$ in dual-unitary circuits, which initially increases the shadow norm more quickly compared to brick-wall Clifford gates, since dual-unitary circuits have maximal butterfly velocity $v_B=1$. Therefore, we look instead at operators with full support, thereby eliminating the ballistic dispersion of the occupation as a factor and directly comparing the rates of saturation of dual-unitary and Clifford brick-wall circuits. Consistent with our intuition that dual-unitary circuits are fast-scrambling, we find that brick-wall dual-unitary circuits can outperform brick-wall Clifford circuits at approximating Clifford measurements and predicting operators with full support.

In \figref{fig:du_sat_rate}, we compare the approach of the base of the shadow norm towards the (global) Clifford measurement limit of $\beta=2$ for the dual-unitary and the Clifford brick-wall cases. We find that the dual-unitary brick-wall more quickly saturates towards $\beta=2$ provided that the circuit depth $t$ is large enough. For $\alpha=2/3$, for operators with full support, the shadow norm and base scaling $\beta$ are uniformly lower than for the Clifford case for all values of $t$. In \figref{fig:du_sat_rate}, we also examine the rate of decay as a function of the circuit depth $t$ for different values of $\alpha$, and compare that against the predictions of the mean-field approximation for $\beta$ in \eqnref{eq:aw2sn}. We see that the functional form is well described by the continuum mean-field approximation, 
\begin{equation}\label{eq:beta-mft-cont}
    \beta-2=\frac{1}{(1+\beta_1) e^{\alpha' (t-1)}-1},
\end{equation}
where $\beta_1 = \frac{1}{3/\sqrt{1+4\alpha/3}-2}$ and $\alpha'$ is the continuum approximation of $\alpha$.

The occupation $\rho^{Cl}(t)$ starting from a uniform weight distribution in the Clifford weight was determined by a similar mean field analysis to scale like \cite{Ippoliti2023O2212.11963},
\begin{equation}
    \rho^{Cl}(t)\sim \frac{c \left(\frac{16}{25}\right)^t}{t^{3/2}}+\frac{3}{4}
\end{equation}
in the long-time limit. This gives the corresponding scaling of the base to be
\begin{equation}
    \beta^{Cl} \sim \frac{1}{1-\frac{2}{3} \rho^{Cl}(t)},
\end{equation}
which decays exponentially with a base of $e^{\log(25/16)}$, compared to $e^{\alpha'}$ in the dual-unitary case. It follows that the dual-unitary should approach the Clifford limit more quickly provided that $\alpha'>\log(25/16)$, which occurs around $\alpha^*_{MF}\approx 0.35$. However, we caution that in practice, it may require a very deep circuit to see the advantage for $\alpha<2/3$.

The ``critical'' value of $\alpha\approx0.35$ makes sense when compared to studies of the rate of Local Operator Entanglement (LOE) growth in brick-wall dual-unitary circuits compared to brick-wall random (Haar) circuits \cite{Bertini2020,Bertini20201}. In these works, the authors investigate linear growth in the LOE as a signature of quantum chaos. In dual-unitary brick-wall circuits, the LOE velocity exceeds the velocity in the Haar (Clifford) case at $\alpha^*\approx 0.32$, which is remarkably close to the $\alpha^*_{MF}$. In appendix B, we show that the saturation rate for $\beta$ in the mean-field approximation is proportional to the local operator entanglement.

To summarize, we've introduced dual-unitary circuit shadow tomography as a more feasible alternative to shadow tomography based on Clifford circuits. The feasibility derives from the simple implementation of two-local dual unitary gates from standard logical operations and local scrambling. Compared to brick-wall Clifford circuits, dual unitaries generally perform worse because they more quickly disperse the Pauli weight of a fixed operator with size $k<n$. Despite this inherent weakness for predicting small operators, dual-unitary gates more quickly relax the Pauli weight and can approach the Clifford measurement limit within the bulk faster. Hence, dual-unitary circuits are advantageous for predicting operators with full or almost support $k\approx n$, because in that case the Pauli weight is already dispersed. The advantage can be observed in \figref{fig:du_sat_rate}. We characterize the advantage in a $n$-independent way by studying the scaling exponent $\beta$ of the shadow norm of a $n$-local Pauli string. We find that this scaling exponent, and therefore the overall sample complexity, is lower for dual-unitary gates with large enough $\alpha$. Furthermore, we find that the dual-unitary gates more quickly approach the Clifford limit $\beta=2$. Combined with previous results from \cite{Akhtar2023S2209.02093}, this suggests a more efficient method for the prediction of general operators with full support. Furthermore, we find that for small $n$, the advantage is actually \textit{amplified} due to the finite size effects, which can be seen in \figref{fig:du_sat_rate} c.

\begin{figure*}[t]
    \centering
    \includegraphics[width=1.99\columnwidth]{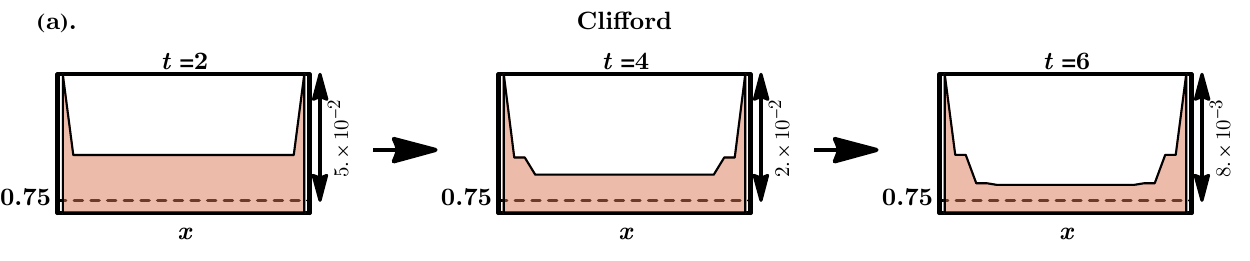}
    \includegraphics[width=1.99\columnwidth]{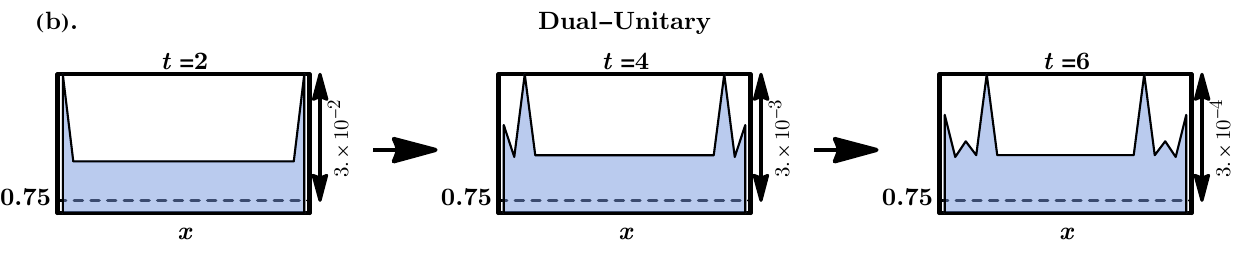}
    \caption{We study the effects of the boundary on the average occupation $\rho(x,t)$ in the Clifford and dual-unitary cases, using numerics from a $n=24$ qubit system for circuit depths $t=2,4,6$, as an illustrative example. (a). In the Clifford case, we find a persistent density of occupation on the boundaries of the system. Because there is no notion of left or right movers in this case, the effect of the boundary is to simply prevent weight from ``leaking'' out, resulting in a higher probability of having operators with support on the boundary. This persistent density of occupation on the boundaries increases the shadow norm according to \eqnref{eq:aw2sn}. In short, the boundary in the brick-wall Clifford case traps excitations, thereby increasing the shadow norm. (b). In the dual-unitary case, there is no persistent density of boundary excitations. This is because excitations in the dual-unitary case are chiral i.e. either left-moving or right-moving. These excitations reflect off the boundary and propagate into the bulk, where the fast-thermalizing nature of the dual-unitary gates causes them to quickly dissipate. Because the boundaries don't trap excitations in the dual-unitary case, the finite size effects are expected to be favorable to the primary claim of this paper. It is also worth noting that the fast-thermalization properties of dual-unitary circuits manifests in the scale of the \(y\)-axis of the figures decaying much faster for the dual-unitary case when compared with the Clifford case.}
    \label{fig:bdry_effects}
\end{figure*}

In the preceding analysis, we've largely ignored the boundary and finite-size effects and correspondingly used the mean-field approximation, which assumes a large system. An important concern could be that such circuits may not be advantageous in near-term devices, where large numbers of qubits cannot be simulated. We now explain why: the boundary effects are in favor of the primary claim of this paper, that the dual-unitary brick-wall circuits are advantageous compared to their Clifford counterparts for predicting operators with extensive support. To understand why, let's explore the role of the boundary in both cases. In the Clifford case, the boundary amounts to the absence of a transfer matrix that could ``leak'' weight outside the system. Because there is no notion of chirality in the Clifford case, when an excitation in the occupation $\rho(x,t)$ reaches the boundary, it simply sits there without diffusing. The build-up of occupation density on the boundaries is actually disadvantageous for prediction, since it slows down the overall thermalization of the system which brings the protocol closer to the global Clifford limit which is known to be advantageous for extensive operators. 

Contrarily, in the dual-unitary case, the excitations must be either left or right-moving. This means that when an excitation in the occupation meets the boundary, it reflects and propagates back into the bulk. Once it enters the bulk, the bulk thermalizing properties of the dual-unitary, which are studied in the mean-field approximation, become dominant. In particular, because the dual-unitaries saturate more quickly in the bulk, the reflected excitations quickly dissipate and the overall weight of the system approaches the global Clifford limit more quickly. Unlike the Clifford case, the boundary excitations, which increase the shadow norm, are forced back into the bulk. This effect is unique to dual-unitary circuits, where there is a notion of chirality, and explains why the boundary does not undermine the primary result of this paper.

We numerically study these boundary effects for a \(n=24\)-qubit system for shallow circuit depths of \(t=2,4,6\) in Figure \ref{fig:bdry_effects}. These numerical simulations highlight a perhaps nontrivial fact that dual-unitary shadow tomography schemes are just as useful for system sizes accessible in the near-term (for operators with full support). Said another way, our scheme is robust to finite-size effects.

\section{Conclusions}

In this work, we explore the average Pauli weight dynamics in dual-unitary brick-wall circuits. We find that it can be described in terms of chiral quasi-particles that propagate at fixed velocities. These particles can fuse into each other and emit one another with a probability determined by the dual-unitary parameter $\alpha$. In particular, at the swap point, $\alpha$ vanishes and the different movers do not interact. We analyze the behavior of the occupation $\rho(x,t)$ as a function of $x,t$ and $\alpha$ and find that for initially contiguous operators it exhibits a ballistic front that moves with maximal light-cone velocity $v_{LC}=1$, and rapidly thermalizes the interior of the light-cone. One advantage of this is that finite depth dual-unitary circuits rapidly approximate the Clifford measurement limit of classical shadow tomography within the bulk, suggesting an advantage for predicting extensive operators.

We focused on qubit systems for simplicity and relevance to classical shadow tomography, but in principle, one could study how varying local Hilbert space dimension affects the results and observations of this paper. Furthermore, the particular brick-wall geometry was also chosen with practical applications in mind, but in principle, one could consider different arrangements of brick-wall gates. Since the brick-wall structure seems essential for the emergence of the chiral quasiparticles, one can ask which of the features of the dual-unitary Pauli weight dynamics are independent of geometry.

In principle, the results of this paper suggest that finite depth, dual-unitary classical shadow tomography could be a powerful tool for predicting extensive operators. We would like for this intuition to be demonstrated on an actual quantum device. Furthermore, it would be helpful to evaluate whether such circuits would be better at fidelity estimation than finite depth Clifford circuits, which were shown to be a more practical alternative to Clifford measurements for predicting low-rank operators \cite{Akhtar2023S2209.02093}.

We use a mean-field treatment instead of an exact treatment of the Pauli weight dynamics. We find that for $\rho(x,t)$, there is practically no difference. However, for calculating the full shadow norm, it is possible that neglecting entanglement for the Pauli weight state could create some discrepancy in our estimation for e.g. the shadow norm of observables with a more complicated spatial support. Therefore, a fuller treatment of the Pauli weight dynamics could be useful.

The emergence of chiral excitations reminiscent of relativistic QFTs in dual-unitary Pauli weight is interesting from the perspective of AdS/CFT correspondence, according to which the entanglement of conformal field theories is related to the geometry of Anti de-Sitter space. Since the Pauli weight is closely related to the second Renyi entropy of the quantum circuit, and since dual-unitary circuits can mimic the entanglement structure in conformal field theories, it stands to reason the emergence of relativity in the Pauli weight dynamics could be a more profound consequence of this correspondence. Can we relate the Pauli weight dynamics to the geometry of AdS space-time? In what other ways does relativity manifest in these circuits?

\section{Acknowledgements}
A.A.A. and Y.Z.Y. are supported by a startup fund from UCSD and the NSF Grant No.~DMR-2238360. N.A. is a KBR employee working under the Prime Contract No. 80ARC020D0010 with the NASA Ames Research Center. J.M. is thankful for support from NASA Academic Mission Services, Contract No. NNA16BD14C. This material is based upon work supported by the U.S. Department of Energy, Office of Science, National Quantum Information Science Research Centers, Superconducting Quantum Materials and Systems Center (SQMS) under contract No. DE-AC02-07CH11359; with this work funded under the NASA-DOE interagency agreement SAA2-403602 governing NASA’s work as part of the SQMS center. The United States Government retains, and by accepting the article for publication, the publisher acknowledges that the United States Government retains, a nonexclusive, paid-up, irrevocable, worldwide license to publish or reproduce the published form of this work or allow others to do so, for United States Government purposes.

\bibliography{ref}

\begin{thebibliography}{10}

\bibitem{Deutsch1991}
J.~M. Deutsch.
\newblock ``Quantum statistical mechanics in a closed system''.
\newblock \href{https://dx.doi.org/10.1103/PhysRevA.43.2046}{Phys. Rev. A {\bf 43}, 2046--2049}~(1991).

\bibitem{Srednicki1994}
Mark Srednicki.
\newblock ``Chaos and quantum thermalization''.
\newblock \href{https://dx.doi.org/10.1103/PhysRevE.50.888}{Phys. Rev. E {\bf 50}, 888--901}~(1994).

\bibitem{Rigol2008}
Marcos Rigol, Vanja Dunjko, and Maxim Olshanii.
\newblock ``Thermalization and its mechanism for generic isolated quantum systems''.
\newblock \href{https://dx.doi.org/10.1038/nature06838}{Nature {\bf 452}, 854--858}~(2008).

\bibitem{Larkin1969QuasiclassicalMI}
Anatoly~I. Larkin and Yu.~N. Ovchinnikov.
\newblock ``Quasiclassical method in the theory of superconductivity''.
\newblock Journal of Experimental and Theoretical Physics~(1969).
\newblock  url:~\url{https://api.semanticscholar.org/CorpusID:117608877}.

\bibitem{Lieb:1972wy}
E.~H. Lieb and D.~W. Robinson.
\newblock ``{The finite group velocity of quantum spin systems}''.
\newblock \href{https://dx.doi.org/10.1007/BF01645779}{Commun. Math. Phys. {\bf 28}, 251--257}~(1972).

\bibitem{D_ra_2017}
Bal{\'{a} }zs D{\'{o}}ra and Roderich Moessner.
\newblock ``Out-of-time-ordered density correlators in luttinger liquids''.
\newblock \href{https://dx.doi.org/10.1103/physrevlett.119.026802}{Physical Review Letters{\bf 119}}~(2017).

\bibitem{Nahum_2017}
Adam Nahum, Jonathan Ruhman, Sagar Vijay, and Jeongwan Haah.
\newblock ``Quantum entanglement growth under random unitary dynamics''.
\newblock \href{https://dx.doi.org/10.1103/physrevx.7.031016}{Physical Review X{\bf 7}}~(2017).

\bibitem{Nahum_2018}
Adam Nahum, Sagar Vijay, and Jeongwan Haah.
\newblock ``Operator spreading in random unitary circuits''.
\newblock \href{https://dx.doi.org/10.1103/physrevx.8.021014}{Physical Review X{\bf 8}}~(2018).

\bibitem{von_Keyserlingk_2018}
C.{\hspace{0.167em} }W. von Keyserlingk, Tibor Rakovszky, Frank Pollmann, and S.{\hspace{0.167em}}L. Sondhi.
\newblock ``Operator hydrodynamics, {OTOCs}, and entanglement growth in systems without conservation laws''.
\newblock \href{https://dx.doi.org/10.1103/physrevx.8.021013}{Physical Review X{\bf 8}}~(2018).

\bibitem{brown2013scrambling}
Winton Brown and Omar Fawzi.
\newblock ``Decoupling with random quantum circuits''.
\newblock \href{https://dx.doi.org/10.1007/s00220-015-2470-1}{Communications in Mathematical Physics {\bf 340}, 867–900}~(2015).
\newblock  \href{http://arxiv.org/abs/1307.0632}{arXiv:1307.0632}.

\bibitem{styliaris_2021_scrambling}
Georgios Styliaris, Namit Anand, and Paolo Zanardi.
\newblock ``Information scrambling over bipartitions: Equilibration, entropy production, and typicality''.
\newblock \href{https://dx.doi.org/10.1103/PhysRevLett.126.030601}{Phys. Rev. Lett. {\bf 126}, 030601}~(2021).

\bibitem{zanardi_2021_openOTOC}
Paolo Zanardi and Namit Anand.
\newblock ``Information scrambling and chaos in open quantum systems''.
\newblock \href{https://dx.doi.org/10.1103/PhysRevA.103.062214}{Phys. Rev. A {\bf 103}, 062214}~(2021).

\bibitem{Nandkishore_2015}
Rahul Nandkishore and David~A. Huse.
\newblock ``Many-body localization and thermalization in quantum statistical mechanics''.
\newblock \href{https://dx.doi.org/10.1146/annurev-conmatphys-031214-014726}{Annual Review of Condensed Matter Physics {\bf 6}, 15--38}~(2015).

\bibitem{Borsi_2022}
M{\'{a} }rton Borsi and Bal{\'{a}}zs Pozsgay.
\newblock ``Construction and the ergodicity properties of dual unitary quantum circuits''.
\newblock \href{https://dx.doi.org/10.1103/physrevb.106.014302}{Physical Review B{\bf 106}}~(2022).

\bibitem{Ho2017E1508.03784}
Wen~Wei {Ho} and Dmitry~A. {Abanin}.
\newblock ``{Entanglement dynamics in quantum many-body systems}''.
\newblock \href{https://dx.doi.org/10.1103/PhysRevB.95.094302}{Physical Review B {\bf 95}, 094302}~(2017).
\newblock  \href{http://arxiv.org/abs/1508.03784}{arXiv:1508.03784}.

\bibitem{Bohrdt2017S1612.02434}
A.~{Bohrdt}, C.~B. {Mendl}, M.~{Endres}, and M.~{Knap}.
\newblock ``{Scrambling and thermalization in a diffusive quantum many-body system}''.
\newblock \href{https://dx.doi.org/10.1088/1367-2630/aa719b}{New Journal of Physics {\bf 19}, 063001}~(2017).
\newblock  \href{http://arxiv.org/abs/1612.02434}{arXiv:1612.02434}.

\bibitem{Nahum2017Q1608.06950}
Adam {Nahum}, Jonathan {Ruhman}, Sagar {Vijay}, and Jeongwan {Haah}.
\newblock ``{Quantum Entanglement Growth under Random Unitary Dynamics}''.
\newblock \href{https://dx.doi.org/10.1103/PhysRevX.7.031016}{Physical Review X {\bf 7}, 031016}~(2017).
\newblock  \href{http://arxiv.org/abs/1608.06950}{arXiv:1608.06950}.

\bibitem{Kukuljan2017W1701.09147}
Ivan {Kukuljan}, Sa{\v{s}}o {Grozdanov}, and Toma{\v{z}} {Prosen}.
\newblock ``{Weak quantum chaos}''.
\newblock \href{https://dx.doi.org/10.1103/PhysRevB.96.060301}{Physical Review B {\bf 96}, 060301}~(2017).
\newblock  \href{http://arxiv.org/abs/1701.09147}{arXiv:1701.09147}.

\bibitem{Nahum2018O1705.08975}
Adam {Nahum}, Sagar {Vijay}, and Jeongwan {Haah}.
\newblock ``{Operator Spreading in Random Unitary Circuits}''.
\newblock \href{https://dx.doi.org/10.1103/PhysRevX.8.021014}{Physical Review X {\bf 8}, 021014}~(2018).
\newblock  \href{http://arxiv.org/abs/1705.08975}{arXiv:1705.08975}.

\bibitem{von-Keyserlingk2018O1705.08910}
C.~W. {von Keyserlingk}, Tibor {Rakovszky}, Frank {Pollmann}, and S.~L. {Sondhi}.
\newblock ``{Operator Hydrodynamics, OTOCs, and Entanglement Growth in Systems without Conservation Laws}''.
\newblock \href{https://dx.doi.org/10.1103/PhysRevX.8.021013}{Physical Review X {\bf 8}, 021013}~(2018).
\newblock  \href{http://arxiv.org/abs/1705.08910}{arXiv:1705.08910}.

\bibitem{Khemani2018O1710.09835}
Vedika {Khemani}, Ashvin {Vishwanath}, and David~A. {Huse}.
\newblock ``{Operator Spreading and the Emergence of Dissipative Hydrodynamics under Unitary Evolution with Conservation Laws}''.
\newblock \href{https://dx.doi.org/10.1103/PhysRevX.8.031057}{Physical Review X {\bf 8}, 031057}~(2018).
\newblock  \href{http://arxiv.org/abs/1710.09835}{arXiv:1710.09835}.

\bibitem{Rakovszky2018D1710.09827}
Tibor {Rakovszky}, Frank {Pollmann}, and C.~W. {von Keyserlingk}.
\newblock ``{Diffusive Hydrodynamics of Out-of-Time-Ordered Correlators with Charge Conservation}''.
\newblock \href{https://dx.doi.org/10.1103/PhysRevX.8.031058}{Physical Review X {\bf 8}, 031058}~(2018).
\newblock  \href{http://arxiv.org/abs/1710.09827}{arXiv:1710.09827}.

\bibitem{Chan2018S1712.06836}
Amos {Chan}, Andrea {De Luca}, and J.~T. {Chalker}.
\newblock ``{Solution of a Minimal Model for Many-Body Quantum Chaos}''.
\newblock \href{https://dx.doi.org/10.1103/PhysRevX.8.041019}{Physical Review X {\bf 8}, 041019}~(2018).
\newblock  \href{http://arxiv.org/abs/1712.06836}{arXiv:1712.06836}.

\bibitem{Zhou2019E1804.09737}
Tianci {Zhou} and Adam {Nahum}.
\newblock ``{Emergent statistical mechanics of entanglement in random unitary circuits}''.
\newblock \href{https://dx.doi.org/10.1103/PhysRevB.99.174205}{Physical Review B {\bf 99}, 174205}~(2019).
\newblock  \href{http://arxiv.org/abs/1804.09737}{arXiv:1804.09737}.

\bibitem{Zhou2019O1805.09307}
Tianci {Zhou} and Xiao {Chen}.
\newblock ``{Operator dynamics in a Brownian quantum circuit}''.
\newblock \href{https://dx.doi.org/10.1103/PhysRevE.99.052212}{Physical Review E {\bf 99}, 052212}~(2019).
\newblock  \href{http://arxiv.org/abs/1805.09307}{arXiv:1805.09307}.

\bibitem{Qi2019M1906.00524}
Xiao-Liang {Qi}, Emily~J. {Davis}, Avikar {Periwal}, and Monika {Schleier-Smith}.
\newblock ``{Measuring operator size growth in quantum quench experiments}''~(2019).
\newblock  \href{http://arxiv.org/abs/1906.00524}{arXiv:1906.00524}.

\bibitem{Xu2019L1805.05376}
Shenglong {Xu} and Brian {Swingle}.
\newblock ``{Locality, Quantum Fluctuations, and Scrambling}''.
\newblock \href{https://dx.doi.org/10.1103/PhysRevX.9.031048}{Physical Review X {\bf 9}, 031048}~(2019).
\newblock  \href{http://arxiv.org/abs/1805.05376}{arXiv:1805.05376}.

\bibitem{Chen2019Q1808.09812}
Xiao {Chen} and Tianci {Zhou}.
\newblock ``{Quantum chaos dynamics in long-range power law interaction systems}''.
\newblock \href{https://dx.doi.org/10.1103/PhysRevB.100.064305}{Physical Review B {\bf 100}, 064305}~(2019).
\newblock  \href{http://arxiv.org/abs/1808.09812}{arXiv:1808.09812}.

\bibitem{Parker2019A1812.08657}
Daniel~E. {Parker}, Xiangyu {Cao}, Alexander {Avdoshkin}, Thomas {Scaffidi}, and Ehud {Altman}.
\newblock ``{A Universal Operator Growth Hypothesis}''.
\newblock \href{https://dx.doi.org/10.1103/PhysRevX.9.041017}{Physical Review X {\bf 9}, 041017}~(2019).
\newblock  \href{http://arxiv.org/abs/1812.08657}{arXiv:1812.08657}.

\bibitem{Kuo2020M1910.11351}
Wei-Ting {Kuo}, A.~A. {Akhtar}, Daniel~P. {Arovas}, and Yi-Zhuang {You}.
\newblock ``{Markovian entanglement dynamics under locally scrambled quantum evolution}''.
\newblock \href{https://dx.doi.org/10.1103/PhysRevB.101.224202}{Physical Review B {\bf 101}, 224202}~(2020).
\newblock  \href{http://arxiv.org/abs/1910.11351}{arXiv:1910.11351}.

\bibitem{Akhtar2020M2006.08797}
A.~A. {Akhtar} and Yi-Zhuang {You}.
\newblock ``{Multiregion entanglement in locally scrambled quantum dynamics}''.
\newblock \href{https://dx.doi.org/10.1103/PhysRevB.102.134203}{Physical Review B {\bf 102}, 134203}~(2020).
\newblock  \href{http://arxiv.org/abs/2006.08797}{arXiv:2006.08797}.

\bibitem{von-Keyserlingk2022O2111.09904}
Curt {von Keyserlingk}, Frank {Pollmann}, and Tibor {Rakovszky}.
\newblock ``{Operator backflow and the classical simulation of quantum transport}''.
\newblock \href{https://dx.doi.org/10.1103/PhysRevB.105.245101}{Physical Review B {\bf 105}, 245101}~(2022).
\newblock  \href{http://arxiv.org/abs/2111.09904}{arXiv:2111.09904}.

\bibitem{Schuster2022O2208.12272}
Thomas Schuster and Norman~Y. Yao.
\newblock ``Operator growth in open quantum systems''.
\newblock \href{https://dx.doi.org/10.1103/PhysRevLett.131.160402}{Phys. Rev. Lett. {\bf 131}, 160402}~(2023).
\newblock  \href{http://arxiv.org/abs/2208.12272}{arXiv:2208.12272}.

\bibitem{Huang2020P2002.08953}
Hsin-Yuan {Huang}, Richard {Kueng}, and John {Preskill}.
\newblock ``{Predicting many properties of a quantum system from very few measurements}''.
\newblock \href{https://dx.doi.org/10.1038/s41567-020-0932-7}{Nature Physics {\bf 16}, 1050--1057}~(2020).
\newblock  \href{http://arxiv.org/abs/2002.08953}{arXiv:2002.08953}.

\bibitem{Ohliger2013E1204.5735}
M.~{Ohliger}, V.~{Nesme}, and J.~{Eisert}.
\newblock ``{Efficient and feasible state tomography of quantum many-body systems}''.
\newblock \href{https://dx.doi.org/10.1088/1367-2630/15/1/015024}{New Journal of Physics {\bf 15}, 015024}~(2013).
\newblock  \href{http://arxiv.org/abs/1204.5735}{arXiv:1204.5735}.

\bibitem{Guta2018F1809.11162}
Madalin {Guta}, Jonas {Kahn}, Richard {Kueng}, and Joel~A. {Tropp}.
\newblock ``{Fast state tomography with optimal error bounds}''.
\newblock \href{https://dx.doi.org/10.1088/1751-8121/ab8111}{Journal of Physics A: Mathematical and Theoretical {\bf 53}, 204001}~(2020).
\newblock  \href{http://arxiv.org/abs/1809.11162}{arXiv:1809.11162}.

\bibitem{Huang2021E2103.07510}
Hsin-Yuan {Huang}, Richard {Kueng}, and John {Preskill}.
\newblock ``{Efficient Estimation of Pauli Observables by Derandomization}''.
\newblock \href{https://dx.doi.org/10.1103/PhysRevLett.127.030503}{Physical Review Letters {\bf 127}, 030503}~(2021).
\newblock  \href{http://arxiv.org/abs/2103.07510}{arXiv:2103.07510}.

\bibitem{Hadfield2020M2006.15788}
Charles {Hadfield}, Sergey {Bravyi}, Rudy {Raymond}, and Antonio {Mezzacapo}.
\newblock ``{Measurements of Quantum Hamiltonians with Locally-Biased Classical Shadows}''.
\newblock Communications in Mathematical Physics {\bf 391}, 951--967~(2022).
\newblock  \href{http://arxiv.org/abs/2006.15788}{arXiv:2006.15788}.

\bibitem{Elben2020M2007.06305}
Andreas {Elben}, Richard {Kueng}, Hsin-Yuan~Robert {Huang}, Rick {van Bijnen}, Christian {Kokail}, Marcello {Dalmonte}, Pasquale {Calabrese}, Barbara {Kraus}, John {Preskill}, Peter {Zoller}, and Beno{\^\i}t {Vermersch}.
\newblock ``{Mixed-State Entanglement from Local Randomized Measurements}''.
\newblock \href{https://dx.doi.org/10.1103/PhysRevLett.125.200501}{Physical Review Letters {\bf 125}, 200501}~(2020).
\newblock  \href{http://arxiv.org/abs/2007.06305}{arXiv:2007.06305}.

\bibitem{Enshan-Koh2022C2011.11580}
Dax {Enshan Koh} and Sabee {Grewal}.
\newblock ``{Classical Shadows With Noise}''.
\newblock \href{https://dx.doi.org/10.48550/arXiv.2011.11580}{{Quantum} {\bf 6}, 776}~(2022).
\newblock  \href{http://arxiv.org/abs/2011.11580}{arXiv:2011.11580}.

\bibitem{Hu2022H2102.10132}
Hong-Ye {Hu} and Yi-Zhuang {You}.
\newblock ``{Hamiltonian-driven shadow tomography of quantum states}''.
\newblock \href{https://dx.doi.org/10.1103/PhysRevResearch.4.013054}{Physical Review Research {\bf 4}, 013054}~(2022).
\newblock  \href{http://arxiv.org/abs/2102.10132}{arXiv:2102.10132}.

\bibitem{Hu2023C2107.04817}
Hong-Ye {Hu}, Soonwon {Choi}, and Yi-Zhuang {You}.
\newblock ``{Classical Shadow Tomography with Locally Scrambled Quantum Dynamics}''.
\newblock \href{https://dx.doi.org/10.1103/PhysRevResearch.5.023027}{Physical Review Research {\bf 5}, 023027}~(2023).
\newblock  \href{http://arxiv.org/abs/2107.04817}{arXiv:2107.04817}.

\bibitem{Levy2021C2110.02965}
Ryan Levy, Di~Luo, and Bryan~K. Clark.
\newblock ``Classical shadows for quantum process tomography on near-term quantum computers''.
\newblock \href{https://dx.doi.org/10.1103/PhysRevResearch.6.013029}{Phys. Rev. Res. {\bf 6}, 013029}~(2024).
\newblock  \href{http://arxiv.org/abs/2110.02965}{arXiv:2110.02965}.

\bibitem{Bu2022C2202.03272}
Kaifeng Bu, Dax~Enshan Koh, Roy~J. Garcia, and Arthur Jaffe.
\newblock ``Classical shadows with pauli-invariant unitary ensembles''.
\newblock \href{https://dx.doi.org/10.1038/s41534-023-00801-w}{npj Quantum Information{\bf 10}}~(2024).
\newblock  \href{http://arxiv.org/abs/2202.03272}{arXiv:2202.03272}.

\bibitem{Hu2022L2203.07263}
Hong-Ye {Hu}, Ryan {LaRose}, Yi-Zhuang {You}, Eleanor {Rieffel}, and Zhihui {Wang}.
\newblock ``{Logical shadow tomography: Efficient estimation of error-mitigated observables}''~(2022).
\newblock  \href{http://arxiv.org/abs/2203.07263}{arXiv:2203.07263}.

\bibitem{Seif2022S2203.07309}
Alireza Seif, Ze-Pei Cian, Sisi Zhou, Senrui Chen, and Liang Jiang.
\newblock ``Shadow distillation: Quantum error mitigation with classical shadows for near-term quantum processors''.
\newblock \href{https://dx.doi.org/10.1103/PRXQuantum.4.010303}{PRX Quantum {\bf 4}, 010303}~(2023).
\newblock  \href{http://arxiv.org/abs/2203.07309}{arXiv:2203.07309}.

\bibitem{Hao-Low2022C2208.08964}
Guang~Hao Low.
\newblock ``Classical shadows of fermions with particle number symmetry''~(2024).
\newblock  \href{http://arxiv.org/abs/2208.08964}{arXiv:2208.08964}.

\bibitem{Akhtar2023S2209.02093}
Ahmed~A. {Akhtar}, Hong-Ye {Hu}, and Yi-Zhuang {You}.
\newblock ``{Scalable and Flexible Classical Shadow Tomography with Tensor Networks}''.
\newblock \href{https://dx.doi.org/10.22331/q-2023-06-01-1026}{Quantum {\bf 7}, 1026}~(2023).
\newblock  \href{http://arxiv.org/abs/2209.02093}{arXiv:2209.02093}.

\bibitem{Bertoni2022S2209.12924}
Christian Bertoni, Jonas Haferkamp, Marcel Hinsche, Marios Ioannou, Jens Eisert, and Hakop Pashayan.
\newblock ``Shallow shadows: Expectation estimation using low-depth random clifford circuits''.
\newblock \href{https://dx.doi.org/10.1103/PhysRevLett.133.020602}{Phys. Rev. Lett. {\bf 133}, 020602}~(2024).
\newblock  \href{http://arxiv.org/abs/2209.12924}{arXiv:2209.12924}.

\bibitem{Arienzo2022C2211.09835}
Mirko Arienzo, Markus Heinrich, Ingo Roth, and Martin Kliesch.
\newblock ``Closed-form analytic expressions for shadow estimation with brickwork circuits''.
\newblock \href{https://dx.doi.org/10.26421/qic23.11-12-5}{Quantum Information and Computation {\bf 23}, 961–993}~(2023).
\newblock  \href{http://arxiv.org/abs/2211.09835}{arXiv:2211.09835}.

\bibitem{Ippoliti2023O2212.11963}
Matteo {Ippoliti}, Yaodong {Li}, Tibor {Rakovszky}, and Vedika {Khemani}.
\newblock ``{Operator Relaxation and the Optimal Depth of Classical Shadows}''.
\newblock \href{https://dx.doi.org/10.1103/PhysRevLett.130.230403}{Physical Review Letters {\bf 130}, 230403}~(2023).
\newblock  \href{http://arxiv.org/abs/2212.11963}{arXiv:2212.11963}.

\bibitem{Hu2024D2402.17911}
Hong-Ye Hu, Andi Gu, Swarnadeep Majumder, Hang Ren, Yipei Zhang, Derek~S. Wang, Yi-Zhuang You, Zlatko Minev, Susanne~F. Yelin, and Alireza Seif.
\newblock ``Demonstration of robust and efficient quantum property learning with shallow shadows''.
\newblock \href{https://dx.doi.org/10.1038/s41467-025-57349-w}{Nature Communications{\bf 16}}~(2025).
\newblock  \href{http://arxiv.org/abs/2402.17911}{arXiv:2402.17911}.

\bibitem{akhtar2023measurementinduced}
Ahmed~A. Akhtar, Hong-Ye Hu, and Yi-Zhuang You.
\newblock ``Measurement-induced criticality is tomographically optimal''.
\newblock \href{https://dx.doi.org/10.1103/PhysRevB.109.094209}{Phys. Rev. B {\bf 109}, 094209}~(2024).
\newblock  \href{http://arxiv.org/abs/2308.01653}{arXiv:2308.01653}.

\bibitem{Bertini2020}
Bruno Bertini, Pavel Kos, and Tomaz Prosen.
\newblock ``{Operator Entanglement in Local Quantum Circuits II: Solitons in Chains of Qubits}''.
\newblock \href{https://dx.doi.org/10.21468/SciPostPhys.8.4.068}{SciPost Phys. {\bf 8}, 068}~(2020).

\bibitem{holdendye2023fundamental}
Tom Holden-Dye, Lluis Masanes, and Arijeet Pal.
\newblock ``Fundamental charges for dual-unitary circuits''.
\newblock \href{https://dx.doi.org/10.22331/q-2025-01-30-1615}{{Quantum} {\bf 9}, 1615}~(2025).
\newblock  \href{http://arxiv.org/abs/2312.14148}{arXiv:2312.14148}.

\bibitem{PhysRevX.9.021033}
Bruno Bertini, Pavel Kos, and Toma\ifmmode \check{z}\else~\v{z}\fi{} Prosen.
\newblock ``Entanglement spreading in a minimal model of maximal many-body quantum chaos''.
\newblock \href{https://dx.doi.org/10.1103/PhysRevX.9.021033}{Phys. Rev. X {\bf 9}, 021033}~(2019).

\bibitem{PhysRevLett.123.210601}
Bruno Bertini, Pavel Kos, and Toma\ifmmode \check{z}\else~\v{z}\fi{} Prosen.
\newblock ``Exact correlation functions for dual-unitary lattice models in $1+1$ dimensions''.
\newblock \href{https://dx.doi.org/10.1103/PhysRevLett.123.210601}{Phys. Rev. Lett. {\bf 123}, 210601}~(2019).

\bibitem{PhysRevB.101.094304}
Lorenzo Piroli, Bruno Bertini, J.~Ignacio Cirac, and Toma\ifmmode \check{z}\else~\v{z}\fi{} Prosen.
\newblock ``Exact dynamics in dual-unitary quantum circuits''.
\newblock \href{https://dx.doi.org/10.1103/PhysRevB.101.094304}{Phys. Rev. B {\bf 101}, 094304}~(2020).

\bibitem{PhysRevLett.126.100603}
Pieter~W. Claeys and Austen Lamacraft.
\newblock ``Ergodic and nonergodic dual-unitary quantum circuits with arbitrary local hilbert space dimension''.
\newblock \href{https://dx.doi.org/10.1103/PhysRevLett.126.100603}{Phys. Rev. Lett. {\bf 126}, 100603}~(2021).

\bibitem{Kos_Styliaris_2023circuitsofspacetime}
Pavel Kos and Georgios Styliaris.
\newblock ``Circuits of space and time quantum channels''.
\newblock \href{https://dx.doi.org/10.22331/q-2023-05-24-1020}{{Quantum} {\bf 7}, 1020}~(2023).

\bibitem{Bensa2021FastestScrambler}
Ja\ifmmode \check{s}\else~\v{s}\fi{} Bensa and Marko \ifmmode \check{Z}\else \v{Z}\fi{}nidari\ifmmode~\check{c}\else \v{c}\fi{}.
\newblock ``Fastest local entanglement scrambler, multistage thermalization, and a non-hermitian phantom''.
\newblock \href{https://dx.doi.org/10.1103/PhysRevX.11.031019}{Phys. Rev. X {\bf 11}, 031019}~(2021).

\bibitem{Cirac2021}
J.~Ignacio Cirac, David P\'erez-Garc\'{\i}a, Norbert Schuch, and Frank Verstraete.
\newblock ``Matrix product states and projected entangled pair states: Concepts, symmetries, theorems''.
\newblock \href{https://dx.doi.org/10.1103/RevModPhys.93.045003}{Rev. Mod. Phys. {\bf 93}, 045003}~(2021).

\bibitem{zanardi_opent_2001}
Paolo Zanardi.
\newblock ``Entanglement of quantum evolutions''.
\newblock \href{https://dx.doi.org/10.1103/PhysRevA.63.040304}{Phys. Rev. A {\bf 63}, 040304}~(2001).

\bibitem{You2018E1803.10425}
Yi-Zhuang {You} and Yingfei {Gu}.
\newblock ``{Entanglement features of random Hamiltonian dynamics}''.
\newblock \href{https://dx.doi.org/10.1103/PhysRevB.98.014309}{Physical Review B {\bf 98}, 014309}~(2018).
\newblock  \href{http://arxiv.org/abs/1803.10425}{arXiv:1803.10425}.

\bibitem{PhysRevLett.125.070501}
Suhail~Ahmad Rather, S.~Aravinda, and Arul Lakshminarayan.
\newblock ``Creating ensembles of dual unitary and maximally entangling quantum evolutions''.
\newblock \href{https://dx.doi.org/10.1103/PhysRevLett.125.070501}{Phys. Rev. Lett. {\bf 125}, 070501}~(2020).

\bibitem{PhysRevA.109.022610}
Shrigyan Brahmachari, Rohan~Narayan Rajmohan, Suhail~Ahmad Rather, and Arul Lakshminarayan.
\newblock ``Dual unitaries as maximizers of the distance to local product gates''.
\newblock \href{https://dx.doi.org/10.1103/PhysRevA.109.022610}{Phys. Rev. A {\bf 109}, 022610}~(2024).

\bibitem{2024arXiv241104898K}
Linghang {Kong}, Zimu {Li}, and Zi-Wen {Liu}.
\newblock ``{Convergence efficiency of quantum gates and circuits}''~(2024).
\newblock  \href{http://arxiv.org/abs/2411.04898}{arXiv:2411.04898}.

\bibitem{2024arXiv241201850W}
Yadong {Wu}, Ce~{Wang}, Juan {Yao}, Hui {Zhai}, Yi-Zhuang {You}, and Pengfei {Zhang}.
\newblock ``{Contractive Unitary and Classical Shadow Tomography}''~(2024).
\newblock  \href{http://arxiv.org/abs/2412.01850}{arXiv:2412.01850}.

\bibitem{ibound}
R.W. Yeung.
\newblock ``A new outlook on shannon's information measures''.
\newblock \href{https://dx.doi.org/10.1109/18.79902}{IEEE Transactions on Information Theory {\bf 37}, 466--474}~(1991).

\bibitem{Farrelly2020reviewofquantum}
Terry Farrelly.
\newblock ``A review of {Q}uantum {C}ellular {A}utomata''.
\newblock \href{https://dx.doi.org/10.22331/q-2020-11-30-368}{{Quantum} {\bf 4}, 368}~(2020).

\bibitem{Grover2015E}
Tarun Grover and Matthew P.~A. Fisher.
\newblock ``Entanglement and the sign structure of quantum states''.
\newblock \href{https://dx.doi.org/10.1103/physreva.92.042308}{Physical Review A{\bf 92}}~(2015).

\bibitem{Bertini20201}
Bruno Bertini, Pavel Kos, and Tomaz Prosen.
\newblock ``{Operator Entanglement in Local Quantum Circuits I: Chaotic Dual-Unitary Circuits}''.
\newblock \href{https://dx.doi.org/10.21468/SciPostPhys.8.4.067}{SciPost Phys. {\bf 8}, 067}~(2020).

\end{thebibliography}

\appendix

\section{Mean-field derivation of the shadow norm bound}
\label{app}

In this section, we wish to prove the bound \eqnref{eq:aw2pw} given by the $n$-th order mean-field approximation. For convenience, we re-produce the bound below in terms of the random binary variables $x=(x_1,\dots x_n)$ where $x_i\in\{0,1\}$, and the probability of a given configuration $p(x)$ is given by the Pauli weight vector $w_{P,t}(A)$, where $i\in A$ ($i\notin A$) if and only if $x_i=1$ ($x_i=0$), prior to the application of the final measurement layer:
\begin{equation}\label{eq:rbv-bound}
    \mathbb{E}[\prod_i (1/3)^{x_i}]\leq \prod_i\mathbb{E}[ (1/3)^{x_i}].
\end{equation}
To see the equivalence of this bound to \eqnref{eq:aw2pw}, observe that the left hand side can be re-written as
\begin{align}
    \mathbb{E}[\prod_i (1/3)^{x_i}]&=\sum_{x} p(x) (1/3)^{\sum_i x_i} \\
    &=\sum_{A\subset \Omega} w_{P,t}(A)(1/3)^{|A|}\\
    &=w_{\mathcal{E}_\sigma}(P)
\end{align}
where $w_{\mathcal{E}_\sigma}(P)$ is a Pauli weight the classical shadow protocol. On the other hand, we can expand $(1/3)^{x_i}=1-2x_i/3$ because these are binary variables, and then using the linearity of expectation we can express the right hand side in terms of simply single-site occupation:
\begin{align}
    \prod_{i=1}^{n}\mathbb{E}[ (1/3)^{x_i}]&=\prod_{i=1}^{n}\mathbb{E}[1-2x_i/3], \\
    &=\prod_{i=1}^{n}(1-2 \rho(x_i)/3).
\end{align}
Let us now assume that $P$ has full support, i.e. $P=\prod_i P_i$ where $P_i\in\{X,Y,Z\}$, and assume periodic boundary conditions so that the weight distribution is also translation invariant. This is expressed in the single-site occupation $\rho=\rho(t)$ having no spatial dependence, with initial boundary condition $\rho(0)=1$. Furthermore, the multi-region occupations $\rho(A)$ only depend on the separation between the sites in $A$. We also assume that the system size $n$ is arbitrarily large. We also assume that the correlations, expressed by the connected, equal-time, two-point correlation function, 
\begin{equation}
    \cov(x_i,x_j):=\mathbb{E}[(x_i-\rho)(x_j-\rho)]=\mathbb{E}[x_i x_j] - \rho^2,
\end{equation}
is small and vanishes outside of the range of correlations generated by the finite-depth circuit i.e. $\cov(x_i,x_j)=0$ when $|i-j|>t$. This is natural to assume since we are evaluating the Pauli weight starting from a disentangled configuration where the weight is uniformly 1 at each site and only applying a depth $t$ brick-wall circuit. Furthermore, this assumption is verified numerically, where we see $\cov(x_i,x_j)$ vanishes outside of a finite window of correlations of size $2t$. Inside the window, it is small and decays exponentially with alternating sign. However, in the following argument we will only assume that $\cov(x_i,x_j)=0$ for $|i-j|>t$ and $|\cov(x_i,x_j)|\leq \varepsilon e^{-b|i-j|}$ where $b>0$ and $\varepsilon>0$ can in principle depend on $\alpha$ and $t$. Let us now define the primary quantity $\beta^{-1}$ that expresses the scaling of the Pauli weight: 
\begin{equation}\label{eq:beta-def}
    \beta^{-1}:=\mathbb{E}[(1/3)^{\sum_i x_i}]^{1/n}.
\end{equation}
Then, the bound reduces to proving that
\begin{equation}\label{eq:appbound}
    \beta^{-1}\leq 1-2\rho/3,
\end{equation}
at the end of the circuit evolution. First consider the case where the spins are uncorrelated. In this case, it is clear that we have equality, since $\beta^{-1}=\mathbb{E}[(1/3)^{x_i}]=1-2\rho/3$. This happens when $t\rightarrow 0$ (and $\rho\rightarrow 1$)  or when $\alpha=0$ so that no correlations are ever generated. The sum $s:=\sum_i x_i$ is itself a random variable that is a sum of independent and identical variables with mean $\rho$. According to the central limit theorem, the distribution of the sum $s$ approaches a normal distribution for large $n$. As we increase $t$ or $\alpha$, we increase the correlation length of the system, but the correlation is assumed to remain small compared to the size of the system. The interactions ``renormalize'' the the random variables, changing the effective size of the unit cell, but nevertheless the variable $s$ will continue to be normally distributed at lowest order in $n$, albeit its shape may differ. This is the content of the generalized central limit theorem for weakly interacting variables, but physically it is reminiscent of Kadanoff-style renormalization. 
The statement can be made more precise by first defining random variables $y_i:=x_i-\rho$ with mean zero. Then according to the central limit theorem, the sum $s_0:=\sum_i y_i$ approaches a normal distribution in the following sense:
\begin{equation}
    s_0 \big/ \sigma\sqrt{n} \rightarrow \mathcal{N}(0,1),
\end{equation}
where $\sigma^2:=\lim_{n\rightarrow \infty}\mathbb{E}[s_0^2]/n$ is the (normalized) variance of the sum, and $\mathcal{N}(0,1)$ is the standard distribution of mean zero and variance one. Then we may re-write the scaling exponent $\beta^{-1}$ in terms of the variance $\sigma^2$:
\begin{align}
    \beta^{-1}
    &=(1/3)^\rho \left(\mathbb{E}[(1/3)^{\sigma\sqrt{n} \frac{s_0}{\sigma\sqrt{n}}}]\right)^{1/n},\\
    &\rightarrow (1/3)^\rho \left( \int_{z\sim\mathcal{N}(0,1)} dz (1/3)^{\sigma \sqrt{n} z} \right)^{1/n},\\
    &=(1/3)^\rho \exp(\frac{\log(3)^2 \sigma^2}{2}),
\end{align}
in the large-$n$ limit. Similarly, this expression may be inverted, giving the variance $\sigma^2$ in terms of $\beta^{-1}$:
\begin{equation}
    \sigma^2=2\log(3^\rho \beta^{-1})/\log(3)^2.
\end{equation}

\begin{figure}[H]
    \centering
    \includegraphics[width=0.65\columnwidth]{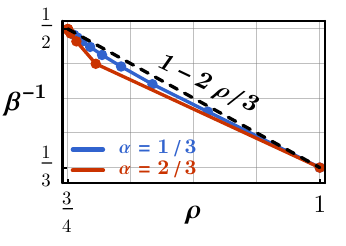}
    \caption{We compare the scaling of the Pauli weight $\beta^{-1}$ for an operator of size $k=n$ in an $n$ qubit system using a finite-depth $t$, brick-wall dual-unitary classical shadow tomography scheme. We eliminate finite-size effects by choosing $n$ to be large, $n=100$, so that we are in the thermodynamic limit. We plot using points $\rho$ at the center vs. $n$-th root of the Pauli weight, for different values of $\alpha=1/3,2/3$. We find that the numerical value for $\beta^{-1}$ sits below the line $1-2\rho/3$ as predicted by the argument above.}
    \label{fig:beta_analysis_1}
\end{figure}

Let us gain some intuition for the behavior of $\sigma^2$. When $t\rightarrow 0$ and $\rho\rightarrow 1$, we know that $\beta^{-1}$ approaches the Pauli measurement limit, $\beta^{-1}\rightarrow 1/3$, so that $\sigma^2\rightarrow 0$ as expected. In this limit, the variables $x_i$ are disentangled. However, as $t\rightarrow \infty$, we know the average weight $\rho$ approaches the saturation value, $\rho\rightarrow 3/4$, and $\beta^{-1}$ approaches the Clifford measurement limit, $\beta^{-1}\rightarrow 1/2$. In this limit, the variance $\sigma^2$ approaches its maximal value $\sigma_\infty^2$, where
\begin{equation}
    \sigma_\infty^2 = 2\left( \frac{\log(1/2)}{\log(3)^2}+\frac{3}{4\log(3)} \right)\approx 0.22.
\end{equation}
We may determine variance $\sigma^2$ in terms of the two-point function. Through minimal assumptions on the two point function, we can show that the bound holds. Let us define the covariance function $f$ as 
\begin{equation}
 f(|i-j|):=\cov(x_i,x_j)=\mathbb{E}[y_i y_j],   
\end{equation}
where we've implicitly assumed translation invariance of the underlying system. Furthermore, $f$ is known to satisfy several properties. For example, we know that $f(0)=\rho(1-\rho)$ since $f(0)$ is simply the variance of a single random variable. Similarly, because the finite depth brick-wall circuit can only generate correlations a fixed distance $t$ away, 
\begin{align}
    \sigma^2&=\lim_{n\rightarrow \infty} \sum_{i,j}\mathbb{E}[y_i y_j]/n, \\
    &=\rho(1-\rho)+2\sum_{i=1}^t f(j),
\end{align}
where the sum over distances $j$ is truncated. Putting this back into the expression for $\beta^{-1}$ gives the scaling factor as a function of $\rho$ (which is itself a function of $\alpha$ and $t$) and the sum of $f(j)$ which also evaluates to a function $\alpha$ and $t$:
\begin{equation}
    \beta^{-1}=(1/3)^\rho \exp(\frac{\log(3)^2}{2}\left( \rho(1-\rho)+2\sum_{i=1}^t f(j) \right)).
\end{equation}
Now, bounding $|f(j)|<\varepsilon e^{-bj}$ gives an upper bound on the variance as 
\begin{equation}
    \sigma^2<\sigma^2_{\max}:=\rho(1-\rho) + (\sigma^2_{\infty}-3/16)(1-e^{-bt}).
\end{equation}
where $t$ is itself a function of $\alpha$ and $\rho$, and $\varepsilon$ has been fixed by requiring that $\sigma^2\rightarrow\sigma^2_{\infty}$ when $t\rightarrow\infty$. What are the properties that $t$ satisfies? Evidently, $t$ is a monotonic positive function of $\alpha$ and $\rho$. It increases from $t=0$ when $\rho=1$ and goes to $t\rightarrow \infty$ when $\rho=3/4$. Furthermore, due to the exponential approach of $\rho$ to 3/4 as $t$ grows large, $t$ is convex in $\rho$ over this interval. The rate at which $t$ interpolates between these values depends on $\alpha$, obviously. When $\alpha$ is large, $\rho$ changes more rapidly. However, it's clear that for any convex, monotonic function $t$ of $\rho$ that goes from $0$ to $\infty$ between the desired limits $\rho=1$ and $\rho=3/4$, the effect of non-zero variance is simply to ``pin'' the value of $\beta^{-1}$ to 1/2 when $\rho\rightarrow 3/4$. 

To finish the argument, observe that substituting $\sigma^2$ for $\sigma^2_{\max}$ upper bounds $\beta^{-1}$ by a convex function, $\beta^{-1}_{\max}$, that touches the line $1-2\rho/3$ at $\rho=1$ and $\rho=3/4$. This is because $\beta^{-1}_{\max}$ is a product of two pieces of convex, decreasing functions. The first piece, $3^{-\rho}\exp(\log(3)^2\rho(1-\rho)/2)$, is convex and decreasing over the interval $\rho\in [3/4,1]$. The second piece, $\exp(\log(3)^2(\sigma^2_{\infty}-3/16)(1-e^{-bt(\rho)}/2)$, is also convex and decreasing. This is because $t(\rho)$ is convex and decreasing. It is straightforward to show that their product must also be convex and decreasing, completing the argument. Therefore, we have that for any $0<\alpha<2/3$, \eqnref{eq:appbound} holds.

We also find that, in practice, $\beta^{-1}$ is close to $1-2\rho/3$. In fact, for discrete time steps, the difference is greatest at $t=1$ with a discrepancy of about 0.02. As $\rho$ approaches $3/4$, the discrepancy goes to zero exponentially fast. This can be seen in the \figref{fig:beta_analysis_1}.

\begin{center}
\noindent\rule[0.5ex]{0.1\linewidth}{0.3pt}\rule[0.5ex]{0.1\linewidth}{0.8pt}\rule[0.5ex]{0.3\linewidth}{1.5pt}\rule[0.5ex]{0.1\linewidth}{0.8pt}\rule[0.5ex]{0.1\linewidth}{0.3pt}
\end{center}

\end{document}